\documentclass[english,twocolumn]{revtex4}
\usepackage[T1]{fontenc}
\usepackage[latin9]{inputenc}
\usepackage{hyperref}
\usepackage[a4paper]{geometry}
\usepackage{graphicx}

\usepackage{babel}

\begin{document}

\title{Dilbert-Peter model of organization effectiveness: computer simulations}

\author{Pawel Sobkowicz}

\email{pawelsobko@gmail.com}

\date{{\small \today}}
\begin{abstract}
We provide a technical report on a computer simulation of general
effectiveness of a hierarchical organization depending on two main
aspects: effects of promotion to managerial levels and efforts to
self-promote of individual employees, reducing their actual productivity.
The combination of judgment by appearance in the promotion to higher
levels of hierarchy and the Peter Principle (which states that people
are promoted to their level of incompetence) results in fast declines
in effectiveness of the organization. The model uses a few synthetic
parameters aimed at reproduction of realistic conditions in typical
multilayer organizations.
\end{abstract}
\maketitle

\section{Introduction}

Computer simulations have become increasingly popular in describing
social phenomena, from traffic jams to opinion formation. The number
of topics and works is very large and there are even popular expositions
of the discipline, such as \citet{ball04-1}. In some cases the new
tools provide significant insight into observed phenomena. Unfortunately,
sometimes the computer simulations and statistical mechanics serve
the same purposes as described by \citet{andreski72-1}, namely to
fortify work that is trivial from sociological point of view. There
is also another pitfall resulting from the focus on the mathematical
and computer tools, namely minimal reference to the actual social
observations. With these dangers in mind it seems quite adventurous
to propose a mathematical model of purely social phenomena, especially
well established in descriptive sociology.

An example of such work, where computer model genuinely allows to
go beyond traditional description, is a recent publication by \citet{pluchino09-1},
who have provided a simple and interesting simulation of the famous
Peter Principle (\citet{peter69-1}). The principle results from observation
that in any hierarchical organization global effectiveness is diminished
due to the fact often skills which make a person an excellent worker
at a given level in the organization may be unsuited at a higher one.
Thus, promotion of the best employees (which looks like a reasonable
way of action) results in loss of skilled workers and, possibly, creation
of less than optimal managers. Of course, such ineffective manager
would no longer be promoted, so eventually everyone would become stuck
at their level of incompetence. Peter originally formulated this idea
in a joke-like fashion, but once we recognize its importance, it becomes
pretty obvious and common sense. There are several works providing
more exact description of the effects of the principle, for example
\citet{kane70-1,fairburn01-1,lazear01-1,lazear04-1}. It should be
noted that companies fight against the {}``incompetence stasis''
resulting from the Peter Principle via constant pressure on employees:
promoting the best is usually accompanied by firing the worst performers
at each level of hierarchy. But while this leads to culling of non-performers,
it might mean loss of capable lower level employees who have been
promoted beyond their skill range. The process creates a bad manager
from a good worker, and then gets rid of the bad manager. This problem
can in turn be solved by providing horizontal career paths, which
ensure that the best specialists could be recognized and awarded without
changing the area of their tasks, so that their skills would not be
lost upon promotion. Yet, despite the fact that Peter Principle is
known for forty years, examples of loss of productivity due to promoting
people beyond their capacities are present in almost all types of
hierarchical organizations, from scientific research to commercial
companies. 

In our model we attempt to go beyond the earlier approaches, which
assume that the basis for promotions is the \emph{actual} performance
of the organization members. Our motivation comes from experience
that decisions to hire and promote are based on \emph{perceived} performance.
Thus, employees or external candidates who focus on presenting themselves
and their own results in good light (instead of just working) have
better chances of advancement. Such individual public relations activity
or \emph{self-promotion} is quite natural and present everywhere:
in commercial companies, universities or political parties. The result
is that not only people reach their incompetency level, but also that
the whole promotion process often focuses on those who have little
to contribute to the organization but rather \emph{spend their effort
and time on getting promoted}. In honor of the Scott Adams' Dilbert
comic strip (\citet{adams96-1}), we propose to call the resulting
process Dilbert-Peter Principle. Similar combination of the two phenomena,
which limit the effectiveness of organizations, has been proposed
by \citet{faria00-1}. It might seem to be presumptuous to use Pointy
Haired Boss and other characters featured in the Dilbert strip as
{}``real world'' basis for computer modelling. Would it not be better
to use official company data, histories and performance figures as
a source for inspiration for a simulation model? In our opinion, just
the contrary. The official documents are usually written with the
purpose of hiding the very mistakes and ineffectiveness we want to
discuss. On the other hand, due to the popularity of the strip and
its WEB site (\url{http://www.dilbert.com/}), the accumulated {}``database''
of cases of stupidity and mismanagement is much better and true to
life than the official corporate profiles and annual reports. The
author's personal experience, gathered in several companies, Polish,
American, German and French, suggests that what looks like jokes is
often an uncannily accurate description modern business organizations.
Thousands of letters sent to Scott Adams from all over the world confirm
this experience. In many areas we find organizations which share similar
activities, sizes and structures, yet which differ widely in their
effectiveness. This is true for both commercial companies and public
organizations. Our goal is to find which which aspects of the model
are crucial in determination of the overall productivity, and perhaps
to suggest measures to improve the operations in real life.

\section{Model description}

The purpose of the model is to provide a simplified description of
effectiveness of a hierarchical organization in which people act to
maximize their promotion chances. The promotions are achieved through
comparisons of productivity of individuals within a given hierarchy
level. One of the key differentiators of the model is the division
between {real productivity} (for example sales results, number
of research publications, lines of code accepted into a project) and
employee \emph{productivity perception} by the authorities responsible
for making promotion decisions. While the effectiveness of the whole
organization is measured by accumulating the \emph{real output} of
the contributors, their advancement within the organization is related
to comparative \emph{perceived results}. Observations show
that often it is not the best performer that gets promoted (as postulated
by the original Peter Principle) but those who appear to be the best
workers. There are thus two main strategies to achieve the promotion.
The first is via improvement of the real output, which corresponds
directly to the Peter model background. The second way, observed all
too frequently, is through the use of political games, thanks to which
a person \emph{appears }to be more productive and worthy of advancement.
These individual PR activities are done at the expense of the actual
work, thereby decreasing the contribution of the individual to the
overall organizational effectiveness. 

Following \citet{pluchino09-1} we propose to compare two possible
scenarios, differing in description of individual productivity\emph{
after} the promotion. The first one, called \emph{continuity scenario}
(`common sense' in \citet{pluchino09-1}), in which the productivity
at the new post is similar to the one at the previous one. Such situation
is expected when the range of tasks related to the new position remains
similar to the previous ones. The \emph{Peter scenario} assumes that
due to difference in tasks at different levels of the organization
the productivity at the new post is unrelated to the old one. In both
cases our current model measures the perceived productivity, including
effects of internal politics. We investigate the effects of promotion
process on key characteristics of the organization: its general
productivity, averages of effectiveness at various levels of the
hierarchy, dynamics of changes due to the promotion preferences. 

The simulations presented here are designed to include several phenomena
associated with modern enterprises and organizations, namely hierarchical
organization, management contribution, measurements of effects dependent
on both individual qualities and cumulative results of subordinates
and interplay between in-company promotion and external hiring. The
goal of the model is to provide some predictions regarding the dependence
of the measured qualities on a few simple controls: importance of
the political gamesmanship for promotion, heritability of skills after
promotion (the continuity model versus Peter hypothesis) and tendency
for internal or external advancement. 

The basic model applies to organizations in which activities are uniform,
i.e. where the nature of tasks is the same for all positions at a
given hierarchy level. Examples might be provided by some government
institutions (such as tax offices), research institutions or by specialized
parts of larger bodies, for example sales divisions in large corporations.
While the basic tasks and measurement criteria per level are comparable
throughout the organization, advancing from one level to another might
change (sometimes very significantly) the nature of the job: from
tax collection or sales to management, more and more remote from the
outside world as we move up in hierarchy

Presented results are based on a small number of selected conditions
and, we hope, can be a starting point for a more advanced study.

\subsection{Hierarchical structure}

We are interested here in department based hierarchical organizations,
such as a commercial company with many local sales offices, or large
software house with multiple project teams. To a certain degree this
description applies to research and government institutions as well. The simplest
way of modeling such organization is via pyramid structure of units
of the same size. Organization is divided into $K$ levels, numbered
from top ($k=1$). Each unit consists of a manager at level $k$ and
$N$ subordinates (workgroup) at level $k+1$. 

The size of the whole organization is thus $N_{TOT}=(N^{K}-1)/(N-1)$.
To achieve reasonable number of levels and unit sizes we may restrict
ourselves to $N\leq10$ and $K=4$ or $K=5$, resulting in organizations
with enough structure but limited to about 1000 people.

\subsection{Agent characteristics}

\subsubsection{Real work and self-promotion\label{sec:realwork}}

Employees are represented by computer agents numbered via subscript
$i$. Each agent is characterized by two parameters. The first, $w_{i}$,
is its capacity to perform effective work at its current position,
called also raw productivity. We consider here two types of contributions:
direct work (for example value of sales achieved by a salesman, lines of code written by a programmer or
research papers published by a scientist) and managerial contribution,
related to organizing, coaching and monitoring efforts of others.
Managers, by their actions, may significantly influence the sum of individual results of their subordinates.
This would correspond, to use our examples, to overall sales figures for a branch office, 
code submitted by a software development group or the research standing of the institute. 
We have decided to combine the two contributions into
one parameter, to keep the spirit of Peter principle, defining one "competence". 

The second characteristics of an agent is the effort
he or she puts into internal political positioning (self-promotion) 
denoted by $p_{i}$. This parameter may include
activities and skills that are quite commonplace in modern organizations,
for example excessive focus on presentation of results compared to the actual
value of one's own results ({}``PowerPoint economy''),
presenting  other's achievements in bad light, forming and nurturing of cliques and
power circles. We consider here that $p_{i}$ is focused on personal
interest of the agent and does not contribute to actual results --
neither those related to direct output, nor to management contributions.
In this work we separate the {}``productive'' part of political activities
(such as organizing, motivating, ensuring cooperation\ldots),
being an inherent part of management activities and contributing to
the real work done, from activities aimed at personal gain.

In our simulations we assume that the $w_{i}$ is drawn from Gaussian
distribution centered around $w_{0}$ (a good value for $w_{0}$ might
be $1$, for reasons explained in Section \ref{sub:Management-contribution})
with distribution width of $\sigma_{w}$, while for $p_{i}$ the distribution
is linear from 0 to $p_{max}$. The two parameters are used to model
the \emph{actual results} of the work done by the agent and the \emph{perception
of the work done} within the organization. The effective contribution
to the organization productivity by the agent is given by\begin{equation}
w'_{i}=w_{i}-p_{i}\end{equation}
This is the simplest way of describing the fact that focusing on improving
one's own position by political activities must deduct some time from the
actual work. On the other hand, effort spent on political maneuvers
improves the perceived results. This is described in a more complex
way, depending on additional parameter: organization's susceptibility
to self-promotion (Eq. \ref{eq:perceived}). Similar trade-off between
technical and social skills has been proposed by \citet{faria00-1}.
It should be noted, however, that in our approach $w_{i}$ contains
\textbf{both} the technical and social skills as defined by Faria.
The change of responsibility upon promotion might result in a different
optimal ratio of the two types of skills, and therefore, in different
value of $w_{i}$ before and after promotion. The treatment of $p_{i}$
is different: it measures an internal characteristic of a person,
its drive to improve own position, which does not improve the actual
results in any way and has not been considered in the past. We have
used values of $w_{0}=1$, $\sigma w=0.3$, $p_{max}=0.6$ in the
simulations. This choice has been motivated by real life observations:
for example such distribution of $w_{i}$ allows differences of work
in the range of 200\% or even more. As for $p_{i}$, the choice of
flat distribution rather than Gaussian reflects relatively frequent
cases of extreme behavior: total lack of focus on self-promotion,
or just the opposite. Author experiences lead to assumption of $p_{max}=0.6$;
there are people who do spend more than half of time on political
ploys aimed to advance their position.

\subsubsection{Management contribution\label{sub:Management-contribution}}

To take into account the fact that in most organizations managers
are measured on the results of the teams they manage, we introduce
here accumulated versions of the effective work and perceived results,
which combine the results of the manager \textbf{and} his or her subordinates.
We have chosen to use a multiplicative way of describing the influence
of the manager on the results of the department. This is by no means
the only choice, but it provides a simple way of describing situation
where a bad manager ($w'_{i}<1$) would actually decrease the overall
output of his department, while the good manager ($w'_{i}>1$) would
increase it. This multiplicative approach differs from additive view
of \citet{pluchino09-1}, who propose that contributions at higher
levels of hierarchy should be modelled by simple addition of manager's
results multiplied by an artificial factor ranging from 0.2 at the
lowest level to 1.0 at the top management, to simulate the increased
importance of the higher levels of hierarchy. Increase of productivity
of given employee over time is a real phenomenon found in many organizations,
related with accumulation of skills and experiences. It would describe
a horizontal progress from an apprentice to a top specialist -- within
a given level of hierarchy. It is less suited to vertical promotion
and managerial tasks, where the contribution of a manager comes from
organizing the work of the subordinates. Moreover, additive approach
can not describe the Dilbertian influence of idiot bosses who often
decrease the results of the departments they manage. The importance
of the individual contribution must be higher as we move up in the
organization levels. Thus we propose that effective results are given
by\begin{equation}
W_{i}=w'_{i}\times\left({\displaystyle \sum_{j\in SUB(i)}}W_{j}\right),\end{equation}
where $SUB(i)$ denotes the agents that are \emph{directly} managed
by agent $i$. Due to recursive nature of the above expression, the
effective results of a manager $i$ include contributions of \emph{all}
its subordinates. For the lowest level of hierarchy $W_{i}=w'_{i}$.
In such model, if all agents have the same value of $w_{i}=1$ and
$p_{i}=0$ (no effort is wasted on political positioning) the results
at a given layer $k$ are $N^{K-k}$, assigning much greater importance
to the managers that in the additive model. But if $w'_{i}<1$ then
the manager decreases the summed contributions of his or hers subordinates.
Such model does not describe well situations where a manager combines
the managerial tasks with the same type of production as his or her
subordinates (e.g. programming team leader writing code, university
department head doing his or her own research or sales manager being directly responsible
for some customers). However, above certain realistic size of the
workgroup, the qualitative results of our model remain unchanged by
such assumption.

\begin{figure*}
\includegraphics[scale=0.5]{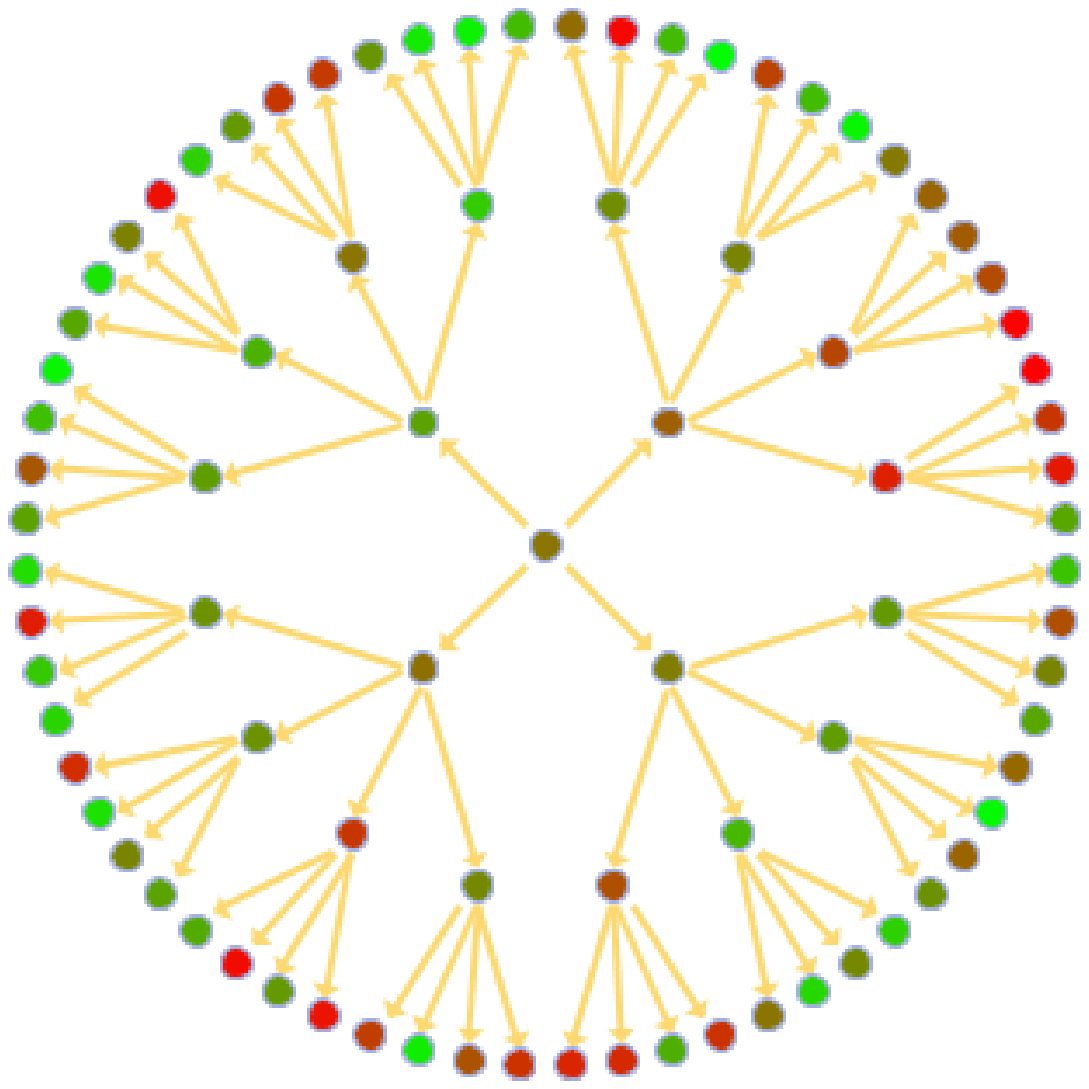} \includegraphics[scale=0.5]{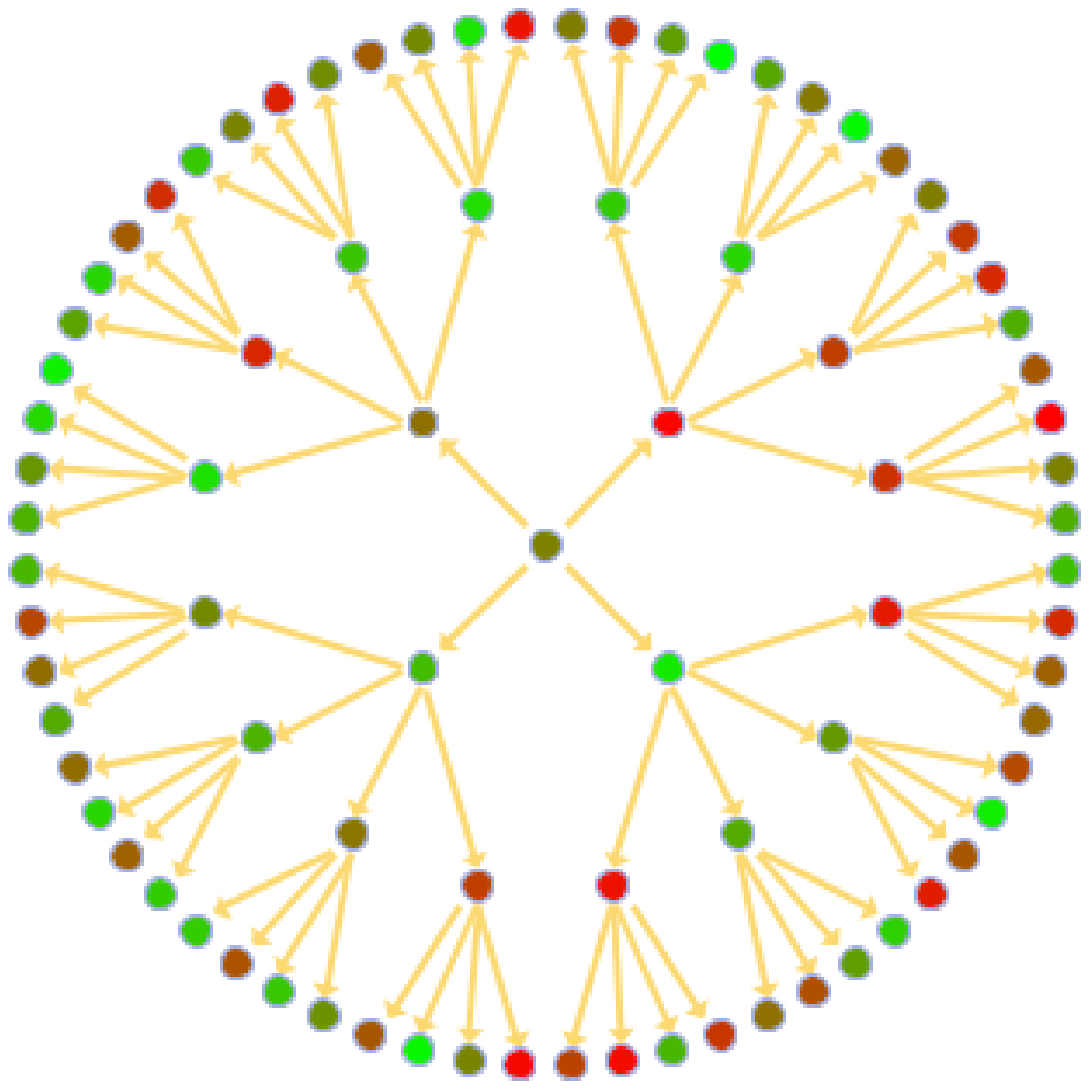}

\caption{Examples of organizations from simulation program. Organization has
4 levels and workgroup size $N$ is also 4. In the left panel colors
correspond to true performance $W_{i}$, in the right panel colors
correspond to perceived performance $U_{i}$; green for highest performers
in given level, red for lowest values, normalized for each level.
\label{Flo:twocircle}}

\end{figure*}

In contrast with the actual results, the {perceived} outcome
for a manager is given by a combination of the real results of his
department (compared to the expected average at the appropriate level)
and the outcome of his political ploys. We have used a simple sum
of these two factors
\begin{equation}
U_{i}=W_{i}/\overline{W(k)}+C\cdot p_{i},\label{eq:perceived}
\end{equation}
where $\overline{W(k)}$ is the average result at the level of the
agent $i$. $C$ is a numerical factor used to model self-promotion
importance - one of the key parameters in the simulations. 
We use the name ``susceptibility'' for $C$, as it determines the relative importance of self-promotion in the selection of candidates for promotion, and thus the way the organization responds to self interests of employees. 
Such form of the perceived results
allows normalization of its two components independently of the level,
number of subordinates etc. Agents with highest and lowest values
of  $U_{i}$ would be the candidates for promotion
and sacking, respectively. Figure \ref{Flo:twocircle} shows sample
organizational structure with  $K=4$ and $N=4$, comparing $W_i$ and $U_i$ at various levels of the organization.
The positions are color coded, with highest
values denoted by green, lowest by red. As may be seen, actual performance
and perceived one may be quite different.

Using the assumptions about the distribution of raw output $w_i$ and self-promotion $p_i$
described in Section~\ref{sec:realwork}, it is possible to calculate the average actual
and perceived results at each organization level in a situation where
the capabilities of each agent are drawn randomly
\begin{equation}
\overline{W(k)}_{RANDOM}=N^{(K-k)}\:\overline{w}_{R}'^{(K-k+1)},\end{equation}
where\begin{equation}
\overline{w}_{R}'=w_{0}-p_{max}/2.\end{equation}

The overall outcome of the political activities on the real and perceived
results of an agent at a given position are simply: increase of $p_{i}$ leads
to decrease of real contribution but increase of the perceived one.
Depending on the value of $C$ playing politics may prove to be advantageous
- or not. As it turns out the result depends on the level of the organization,
making the model more life-like. 

\subsection{Promotion, firing and hiring}

To model the processes of promotion in an organization we need to
provide some assumptions as to the personnel mobility within it. This
requires balancing between model simplicity and the need to cover
realistic situations. The rules proposed apply to organization levels
$k\geq2$. For the top level, there is no competition and no hiring/firing.
Real organizations certainly experience changes at the CEO level,
and such changes can bring large differences in the overall performance.
Charismatic leaders, like Steve Jobs, by their vision and management
skills can radically change the evolution of their companies. Such
contributions are, however, impossible to simulate in a statistical
computer model. Our goal is to model internal competition and promotion
mechanisms within an organization, coming from lower levels. To allow
such focus in our simulations we have assumed that the top manager
has no influence on the overall productivity, having $w_{1}=1$, and
$p_{1}=0$. 

At lower levels the firing and promotion scheme easier to simulate.
At the end of every quarter there is some chance (given by parameter
$x_{f}$) that from each workgroup of $N$ employees subordinate to
a given manager one would be fired. Thus the typical churn rate is
the same at every level, being given by $x_{f}/N$, and the reason
for introducing $x_{f}$ is to allow adjustment of the rate of change
in the model. This would allow the {}``simulation time'' to reflect
actual time for real organizations. The choice of the agent to be
fired in each workgroup is deterministic: the worst perceived performer
is selected as the candidate for firing. 

We have also considered a variant in which a manager threatened by
a possibility of being fired, if sufficiently skilled in political
games, might pass the blame to one of his or her direct reports. In
this variant, the existence of scapegoats further improves the chances
of those agents who focus on politics instead of the real work. However,
such blame-shifting requires high political skills, so we have assumed
that it would be possible only with probability proportional the political
skills of the threatened manager, $p_{i}$. For low values of $p_{i}$
putting the blame to others would be virtually impossible, but for
those who have high values pf $p_{i}$ closer to 1, the security of
the manager in question would be greatly improved, at the expense
of his subordinates. So, in blame-shifting scenario, survival role
of $p_{i}$ is enhanced.

The vacant positions can be filled either by internal promotion or
by external hiring. This can by described by assuming two mechanisms
of filling a given empty post, each with probability being a part
of the model. First, there is a probability that the post will be
filled by externally hired agent (probability $x_{e}$). To keep the
model close to the real hiring process we assume that for each vacant
position some applicants with randomly drawn values of $w_{i}$ and
$p_{i}$would be compared (we used 4 candidates in the simulations).
But it is important to remember that this process would compare the
\emph{perceived} qualities of the candidates. The same political skills
that are used during employment may be a rough measure of how the
candidate presents his or her past achievements and value for the
future employer. So, the hired agent would be the one with the highest
value of $w'_{i}/(w_{0}-p_{max}/2)+C\cdot p_{i}$. It seems a reasonable
assumption that the same susceptibility factor $C$ that is used to
describe the organization's susceptibility to political maneuvering
of the employees would be used for appraisal of candidates. We have
also run simulations where no selection of external candidates has
been performed and thus the new agent would simply have a random set
of $w_{i}$, $p_{i}$ values. This allows to study the effects of pre-selection of external candidates. 

The second option of filling a vacancy is through direct promotion
of the best performer (again, we measure perceived performance!) of
the organizational level directly below the vacancy. 
It is possible to consider two options: local promotion, where the promoted agent is selected from direct
subordinates under the vacated position, and global where anyone from
the lower level can be picked to fill the vacated post. In the preliminary
results presented here the global model has been used. The promotion will, of course,
leave another vacancy to be filled at a lower level. The hiring process
should start at the highest vacancy and continue down. All the vacancies
at the lowest level are, by their nature, filled by external hiring.

To make the model comparable to previous `Peter Principle' studies
two options of treating the agent capacities after a promotion are
possible. They are: the \textbf{continuity hypothesis}, in which the
productivity at the higher level remains close to the previous one
\begin{equation}
w_{i}^{+}=w_{i}+\delta w,\end{equation}
where $w_{i}^{+}$ denotes agents capacity on the higher level. Agent's
productivity changes by a small random value of $\delta w$, with
normal distribution centered at zero with width of $\sigma_{\delta w}$.
This hypothesis would be applicable to situations where the tasks
at the higher hierarchy level are somewhat similar to those at a lower
level (advancement from programmer to programming team leader, for
example). On the other hand, in situations where the new post calls
for totally different set of skills (salesman promoted to sales manager
or to marketing manager position) it is sensible to assume the \textbf{Peter
hypothesis} in which $w_{i}^{+}$ is randomly drawn, without any relation
to previous $w_{i}$. For this reason the promotion modes (local or
global) and the promotion hypotheses (common sense or Peter hypothesis)
might be correlated.

Depending on the organization the change of the scope of activity
due to promotion, from direct production (writing the code, visiting
customers, conducting experiments) to managerial tasks of organizing,
planning and supervision may be more or less gradual. The first promotion,
turning a worker into first line manager introduces the greatest change
and one could expect that Peter hypothesis to hold. On the other hand
further promotions may involve the use of more or less the same set
of managerial skills applied to larger groups and responsibilities,
which would be better described by the common sense approach. It might
make sense then to introduce a \textbf{mixed model} corresponding
to such situation.

\subsection{Simulation considerations}

During simulation process we are looking both at global changes of
organization effectiveness depending on the promotion model as well
as individual career paths and results of individual strategies (given
by unchanging values of the political factor $p_{i}$ of each agent).
The simulation steps should correspond to realistic conditions, e.g.
quarters or semesters -- periods where typically performance of employees
is reviewed. Thus we would be interested in, say, up to 64 steps (16
years) -- but many of the interesting phenomena could happen on the
shorter timescales. By experience, modern commercial organizations
stay in the same shape (without major reorganizations) for periods
of 3-5 years. And, of course, major reorganizations are not covered
by the simplistic model described above. The individual results of
the simulation runs differed significantly (as discussed in Section
\ref{sub:Simulation-statistics}) and we for each combination of parameters
we have accumulated results of 8000 runs, to obtain averages and distribution
of key characteristics.

The key parameters and characteristics of the simulated organization
were:
\begin{itemize}
\item Overall performance of the organization, given by the averages of
effective total result of the topmost boss $W_{TOT}=W_{1}$ and its
changes during simulation steps, as well as effective performances
of organization units at different levels, as given by their bosses'
$W_{i}$.
We also recorded the number of new hires  compared to internal promotions, to check the initial
assumptions It is worth noting that these parameters are comparable to real world data,
ans such comparison could lead to improvements
of the model (for example probabilities of external hiring depending
on the level of the vacancy).
\item Monitoring true productivity and self-promotion factor at various levels of the organization.
Average $\left\langle p_{i}\right\rangle _{k}$ for each level $k$
as well as the average value of raw individual
productiveness $\left\langle w_{i}\right\rangle _{k}$ were calculated at each step of the simulations; 
to check if it is the political manipulators or the real workers who move
to the upper echelons, as suggested by common sense. 
\end{itemize}

\section{Results and discussion}

The model presented in this work is rather complex, involving several
parameters designed to mimic at least some of the crucial aspects
of modern organizational life. This complexity, however, makes deriving
clear dependencies rather difficult. For the purpose of this preliminary
paper we decided to divide the system controls into two groups. The
first contains those that are static between various simulations,
for example the distributions of $w_{i}$ and $p_{i}$.
For most of the simulations we have also kept the firing rate $x_{f}$
and the external hiring rate $x_{e}$ fixed. The values
of these parameters were selected to keep the {}``simulation time'' as close
as possible to the real world, and obtain reasonable values of the
churn ratio and average time spent as given position. 

The second group contained  controls that were varied between simulated organizations, 
describing their main
characteristics: number of levels and workgroup size, 
presence or absence of blame-shifting, pere-selection sample size for external candidates 
and, of course, type of
post-promotion efficiency model (Peter hypothesis, continuity model) and
organization susceptibility to self-promotion.

We present here results for two hierarchies, the first comprised of 5 levels of
5 people in a workgroup, the second of only 4 levels but 9 people
in a workgroup. This choice was motivated by a desire to keep the
overall sizes of the two configurations similar (781 in the first
case, 820 in the second). Also, the number of workers in the lowest
(non-managerial) level was of the same order (625 vs. 729). Based
on the distribution of $w_{i}$ and $p_{i}$ defined in Section 2,
we can derive two {}``yardstick'' measures of the total organization
productivity. The first, which we would call {}``\emph{neutral productivity}'',
assumes that no self-promotion take place ($p_i=0$) and that all workers
and managers have the same effective $w'_{i}=w_i=1$. Then the organization's
output is simply given by the number of lowest level workers (as managers
neither improve nor diminish the results). The other standard is the
average productivity in a wholly random organization, this time including
the negative effects of time and effort lost on political games. The
values of such static, random distribution are, with our choice of
parameters, rather low, (105 for the 5-by-5 case and 175 for the 4-by-9
one). This is, of course, due to the largely negative impact of the
managerial structure, where every manager \emph{decreases} the production
of his/her department by a factor of $w_{R}'=w_{0}-p_{max}/2$ (0.7 in our case).
These values are also starting points of the results of the dynamical
simulations, as the starting conditions used random distribution of
agent characteristics. It should be noted that the random configuration
and associated productivity seems highly unrealistic, as it means
that all managers negatively influence the outcome. For this reason,
we have used the neutral productivity as the criterion of the improvement
vs. decline of the organization effectiveness in the simulations.
Interestingly enough, some combinations of parameters resulted in
configurations that were worse than the random one -- negative selection
is real, especially if the organization itself is under no threat
of existence nor competition, as is often true for government structures. 

\begin{figure*}
\includegraphics[scale=0.8]{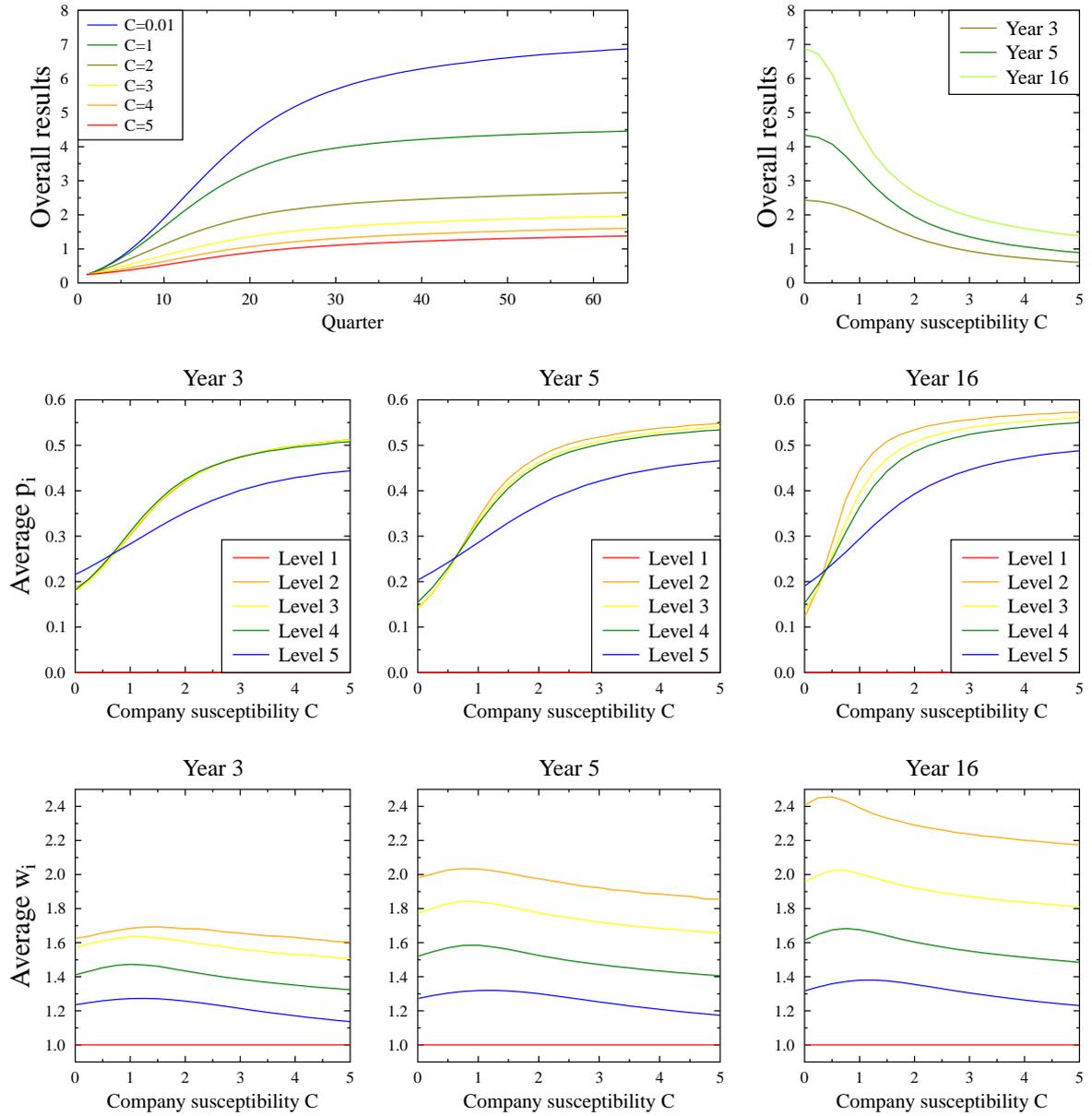}

\caption{Simulation results for the continuity model, 5 layers, 5 positions
in a workgroup. Pre-selection from 4 external candidates (4NH) and
blame-shift process present. Upper panels: time evolution and dependence
of overall productivity $W_{TOT}$ on organization susceptibility
$C$, divided by productivity of \textquotedbl{}neutral\textquotedbl{}
organization of the same size. Lower panels: dependence of average
$p_{i}$ and $w_{i}$ values on $C$ at the end of the third, fifth
and 16th year of evolution. \label{Flo:con5x54nhbl}}

\end{figure*}

\begin{figure*}
\includegraphics[scale=0.8]{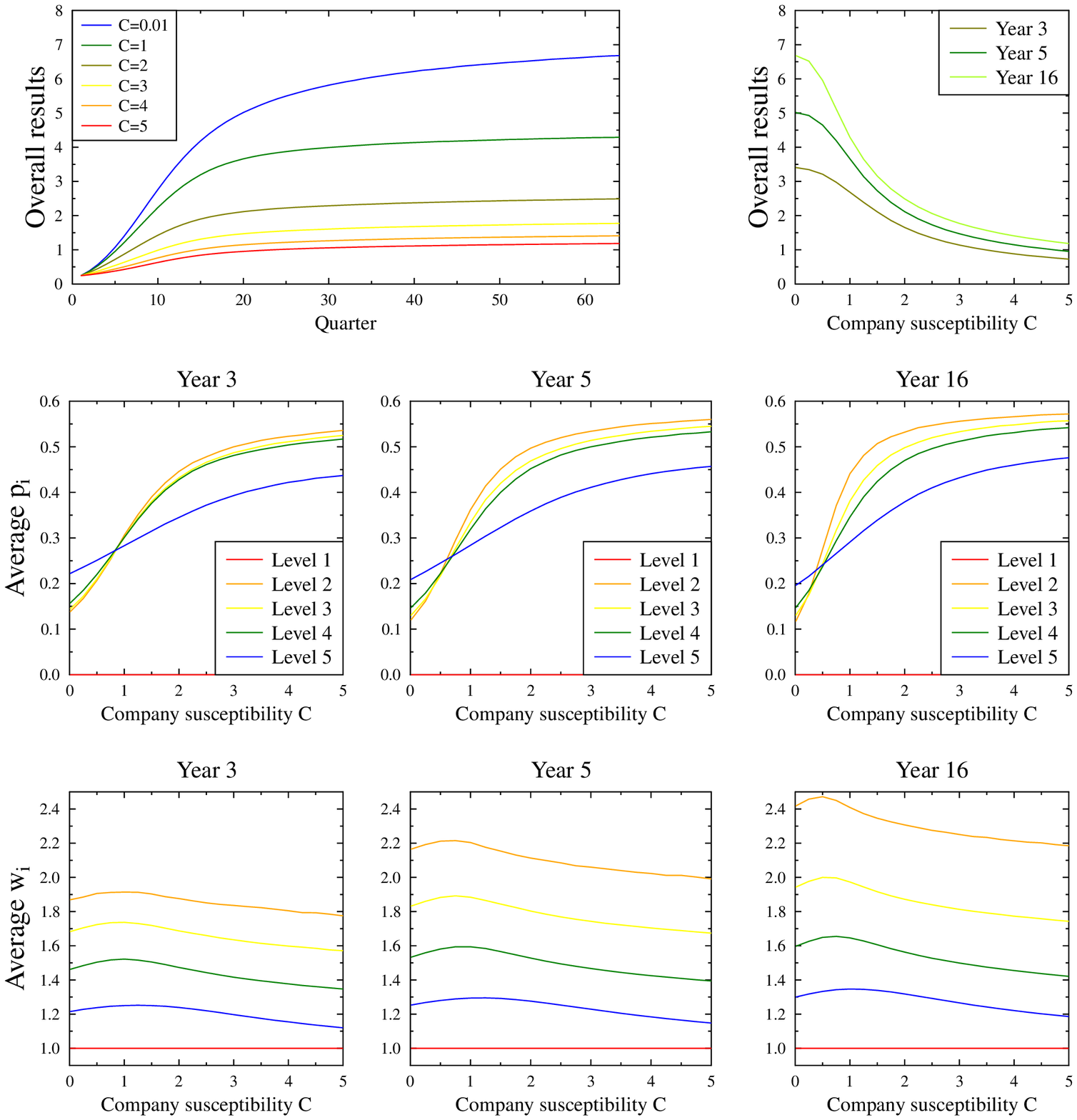}

\caption{Simulation results for the continuity model, 5 layers, 5 positions
in a workgroup. Pre-selection from 4 external candidates (4NH) and
no blame-shift process present. Upper panels: time evolution and dependence
of overall productivity $W_{TOT}$ on organization susceptibility
$C$, divided by productivity of \textquotedbl{}neutral\textquotedbl{}
organization of the same size. Lower panels: dependence of average
$p_{i}$ and $w_{i}$ values on $C$ at the end of the third, fifth
and 16th year of evolution. \label{Flo:con5x54nhnbl}}

\end{figure*}

\begin{figure*}
\includegraphics[scale=0.8]{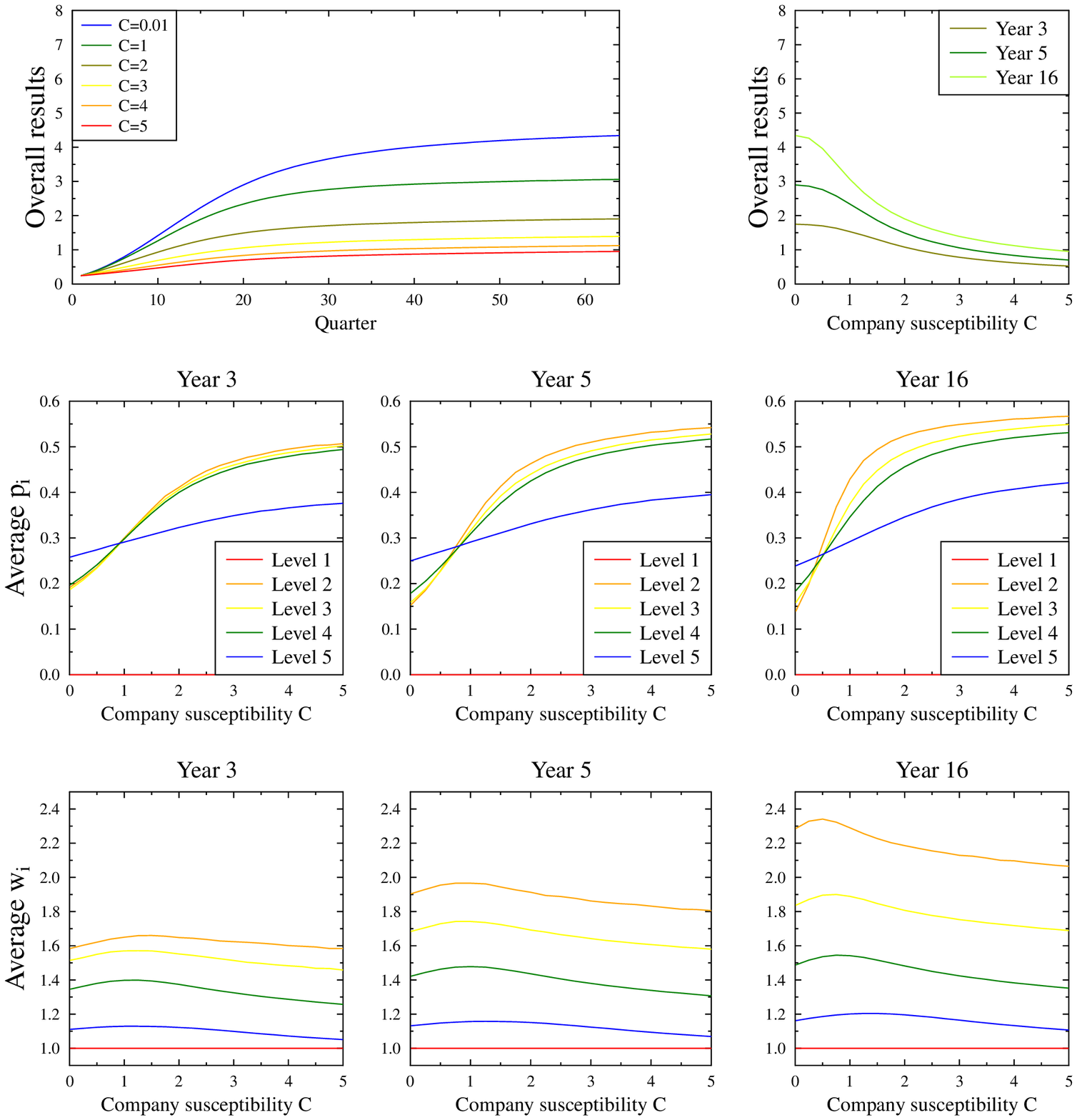}

\caption{Simulation results for the continuity model, 5 layers, 5 positions
in a workgroup. No pre-selection for external candidates (1NH) and
blame-shift process present. Upper panels: time evolution and dependence
of overall productivity $W_{TOT}$ on organization susceptibility
$C$, divided by productivity of \textquotedbl{}neutral\textquotedbl{}
organization of the same size. Lower panels: dependence of average
$p_{i}$ and $w_{i}$ values on $C$ at the end of the third, fifth
and 16th year of evolution. \label{Flo:con5x51nhbl}}

\end{figure*}

\begin{figure*}
\includegraphics[scale=0.8]{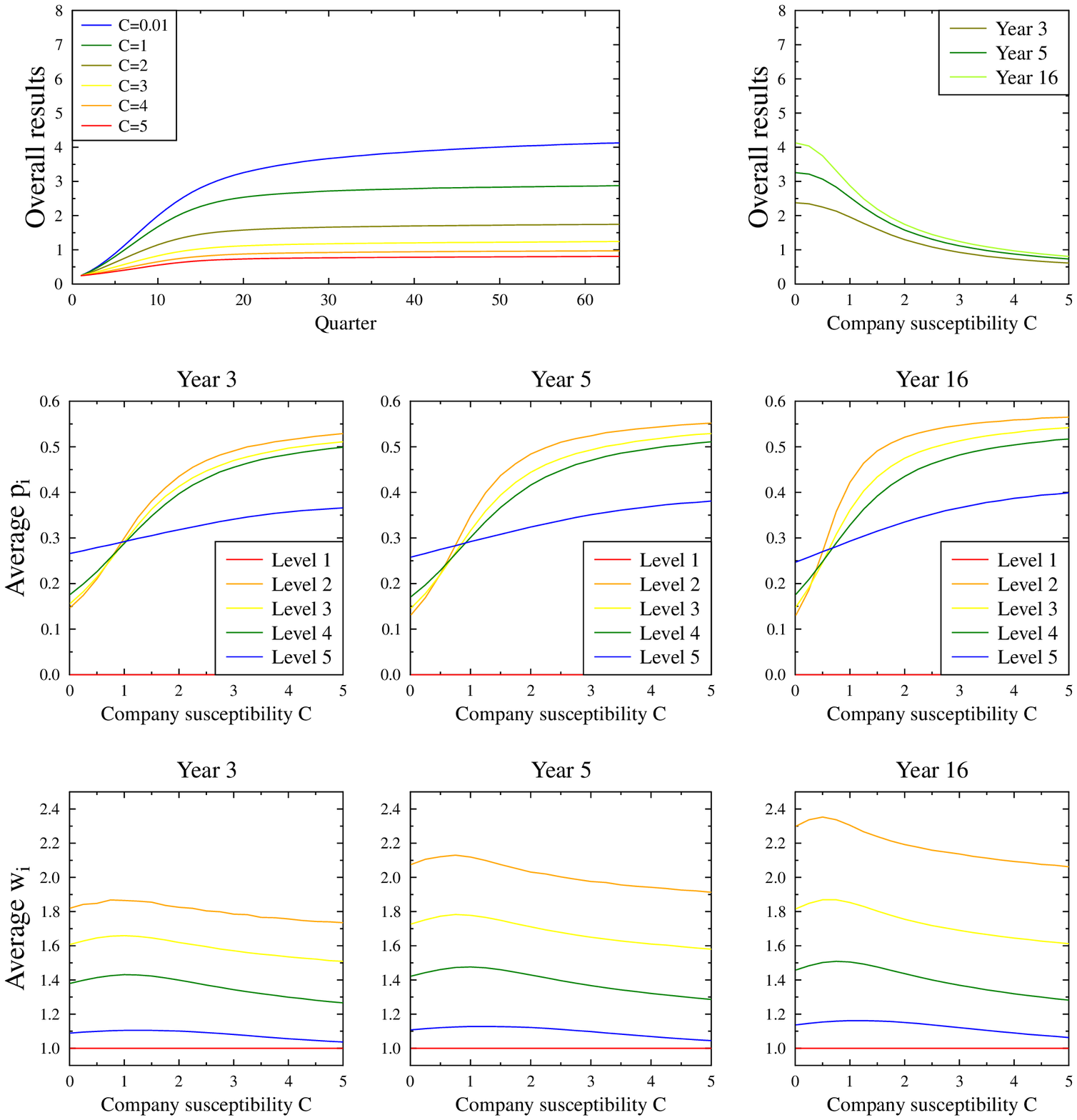}

\caption{Simulation results for the continuity model, 5 layers, 5 positions
in a workgroup. No pre-selection for external candidates (1NH) and
no blame-shift process present. Upper panels: time evolution and dependence
of overall productivity $W_{TOT}$ on organization susceptibility
$C$, divided by productivity of \textquotedbl{}neutral\textquotedbl{}
organization of the same size. Lower panels: dependence of average
$p_{i}$ and $w_{i}$ values on $C$ at the end of the third, fifth
and 16th year of evolution. \label{Flo:con5x51nhnbl}}

\end{figure*}

\begin{figure*}
\includegraphics[scale=0.8]{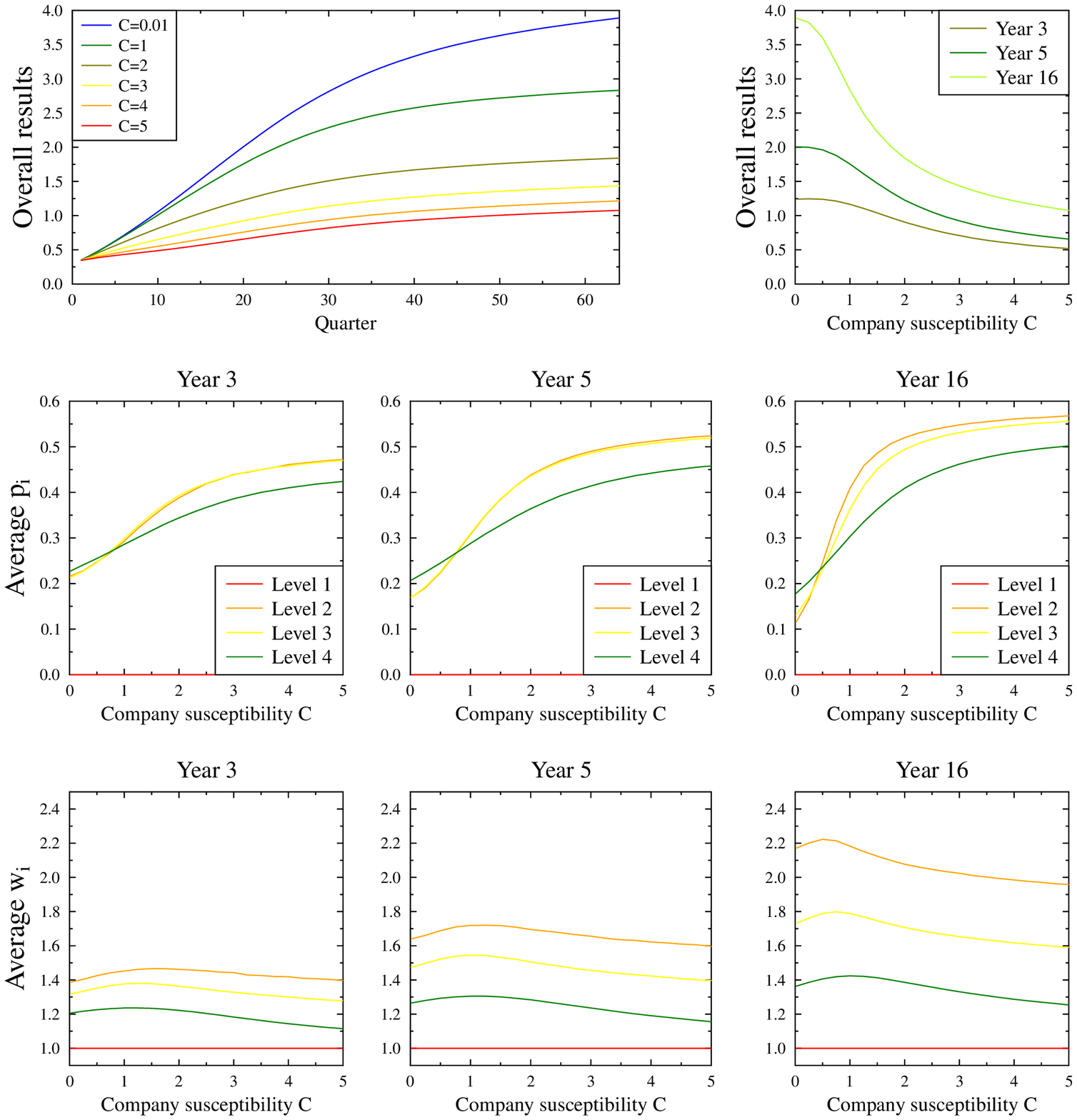}

\caption{Simulation results for the continuity model, 4 layers, 9 positions
in a workgroup. Pre-selection from 4 external candidates (4NH) and
blame-shift process present. Upper panels: time evolution and dependence
of overall productivity $W_{TOT}$ on organization susceptibility
$C$, divided by productivity of \textquotedbl{}neutral\textquotedbl{}
organization of the same size. Lower panels: dependence of average
$p_{i}$ and $w_{i}$ values on $C$ at the end of the third, fifth
and 16th year of evolution. \label{Flo:con4x94nhbl}}

\end{figure*}

\begin{figure*}
\includegraphics[scale=0.8]{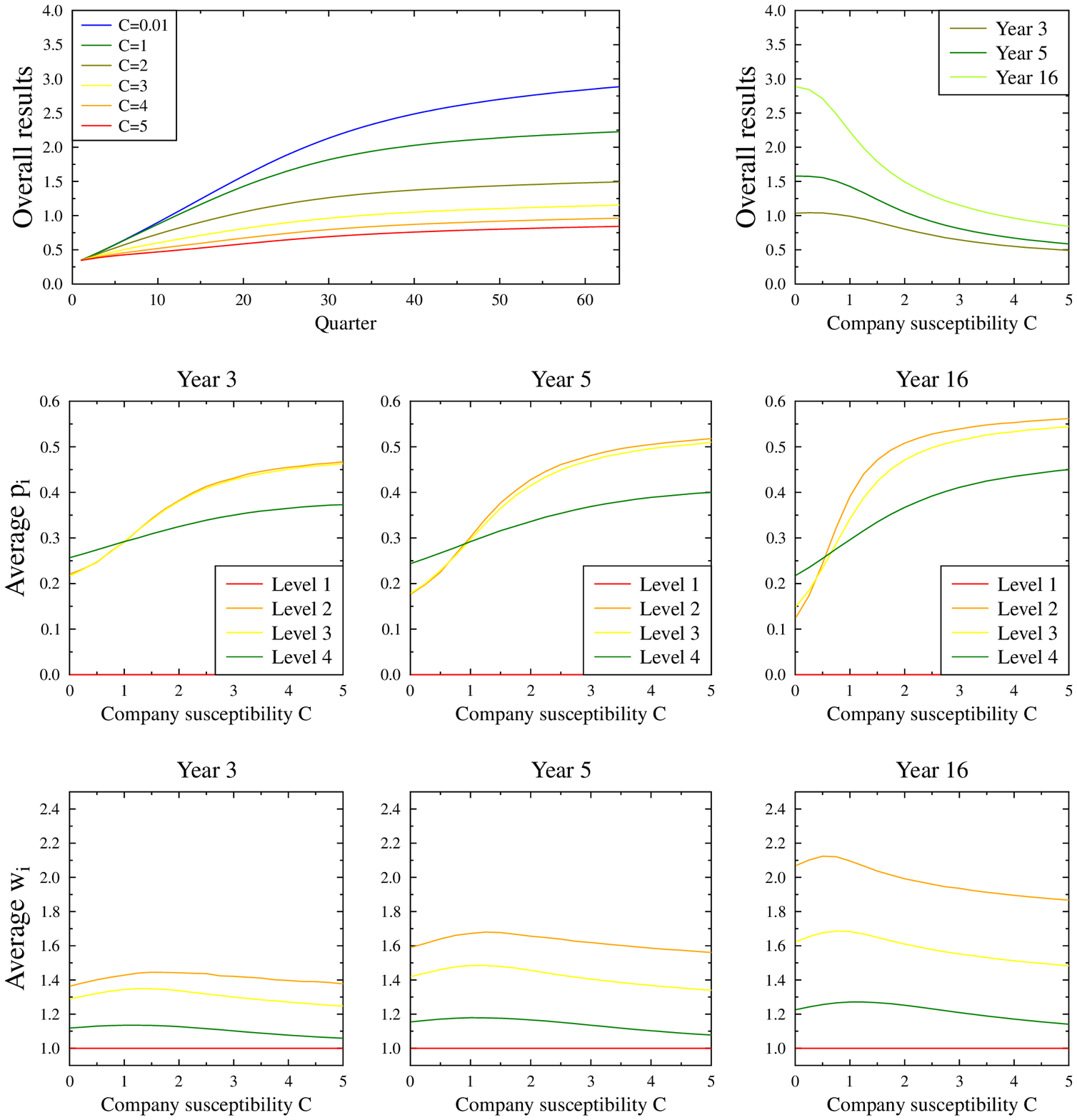}

\caption{Simulation results for the continuity model, 4 layers, 9 positions
in a workgroup. No pre-selection for external candidates (1NH) and
blame-shift process present. Upper panels: time evolution and dependence
of overall productivity $W_{TOT}$ on organization susceptibility
$C$, divided by productivity of \textquotedbl{}neutral\textquotedbl{}
organization of the same size. Lower panels: dependence of average
$p_{i}$ and $w_{i}$ values on $C$ at the end of the third, fifth
and 16th year of evolution. \label{Flo:con4x91nhbl}}

\end{figure*}

\begin{figure*}
\includegraphics[scale=0.8]{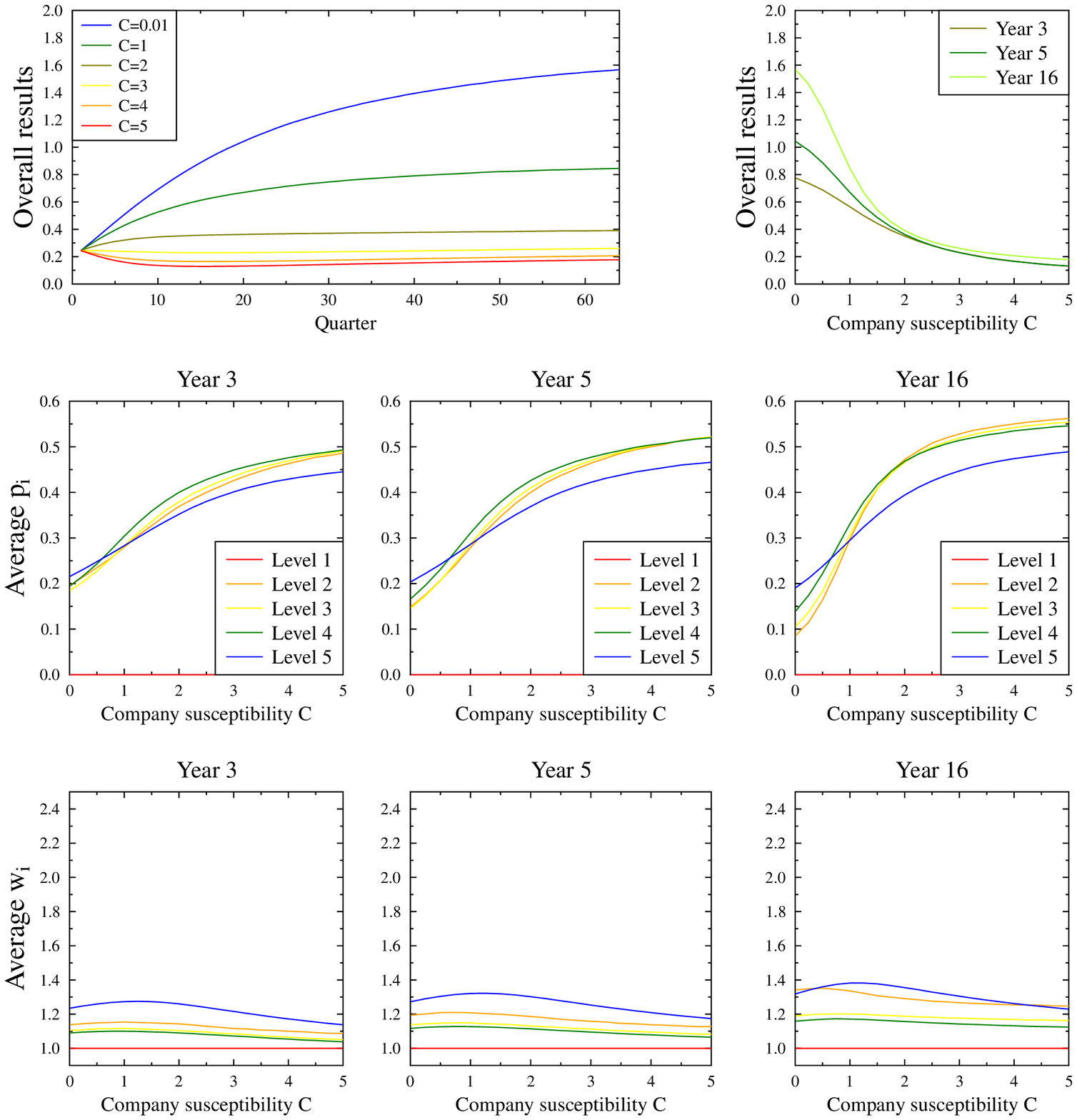}

\caption{Simulation results for the Peter model, 5 layers, 5 positions in a
workgroup. Pre-selection from 4 external candidates (4NH) and blame-shift
process present. Upper panels: time evolution and dependence of overall
productivity $W_{TOT}$ on organization susceptibility $C$, divided
by productivity of \textquotedbl{}neutral\textquotedbl{} organization
of the same size. Lower panels: dependence of average $p_{i}$ and
$w_{i}$ values on $C$ at the end of the third, fifth and 16th year
of evolution. \label{Flo:pet5x54nhbl}}

\end{figure*}

\begin{figure*}
\includegraphics[scale=0.8]{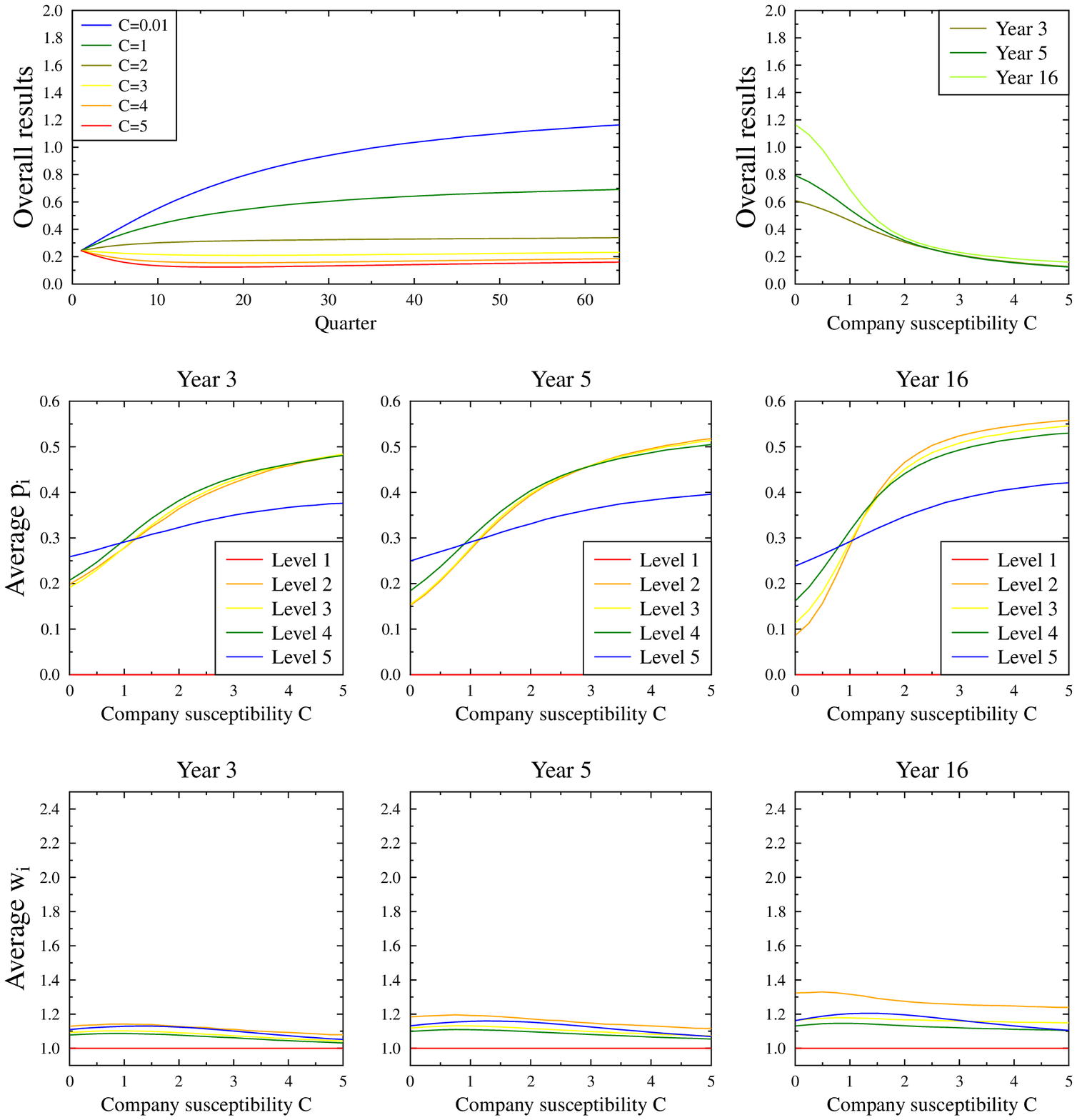}

\caption{Simulation results for the Peter model, 5 layers, 5 positions in a
workgroup. No pre-selection of external candidates (1NH) and blame-shift
process present. Upper panels: time evolution and dependence of overall
productivity $W_{TOT}$ on organization susceptibility $C$, divided
by productivity of \textquotedbl{}neutral\textquotedbl{} organization
of the same size. Lower panels: dependence of average $p_{i}$ and
$w_{i}$ values on $C$ at the end of the third, fifth and 16th year
of evolution. \label{Flo:con5x54nhbl-7}}

\end{figure*}

\begin{figure*}
\includegraphics[scale=0.8]{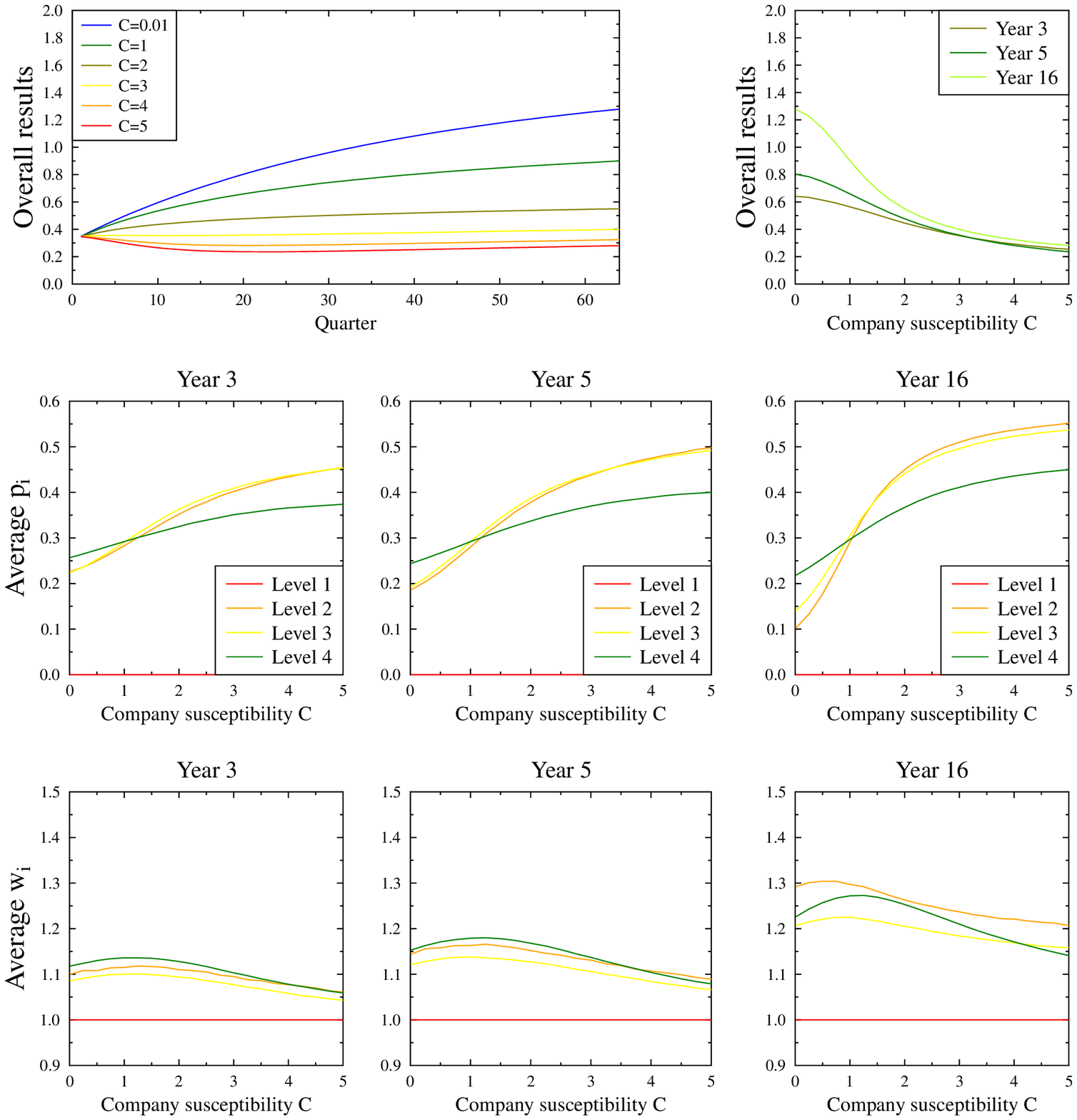}

\caption{Simulation results for the Peter model, 4 layers, 9 positions in a
workgroup. No pre-selection for external candidates (1NH) and blame-shift
process present. Upper panels: time evolution and dependence of overall
productivity $W_{TOT}$ on organization susceptibility $C$, divided
by productivity of \textquotedbl{}neutral\textquotedbl{} organization
of the same size. Lower panels: dependence of average $p_{i}$ and
$w_{i}$ values on $C$ at the end of the third, fifth and 16th year
of evolution. \label{Flo:pet4x91nhbl-1}}

\end{figure*}

Despite the fact that in all simulations the pool of agents' individual
capabilities remained the same, the final long term results of the
model organizations varied by almost an order of
magnitude! At the end of this report we include detailed average
results of the simulations for several combinations of model parameters
(Figures \ref{Flo:con5x54nhbl} to \ref{Flo:pet4x91nhbl-1}). Due to
the interplay of the processes introduced in our model, separation
of the individual contributions is not easy and the impact of various
model parameters requires some detailed analysis.

\subsection{Simulation statistics\label{sub:Simulation-statistics}}

\begin{figure*}
\includegraphics[scale=0.8]{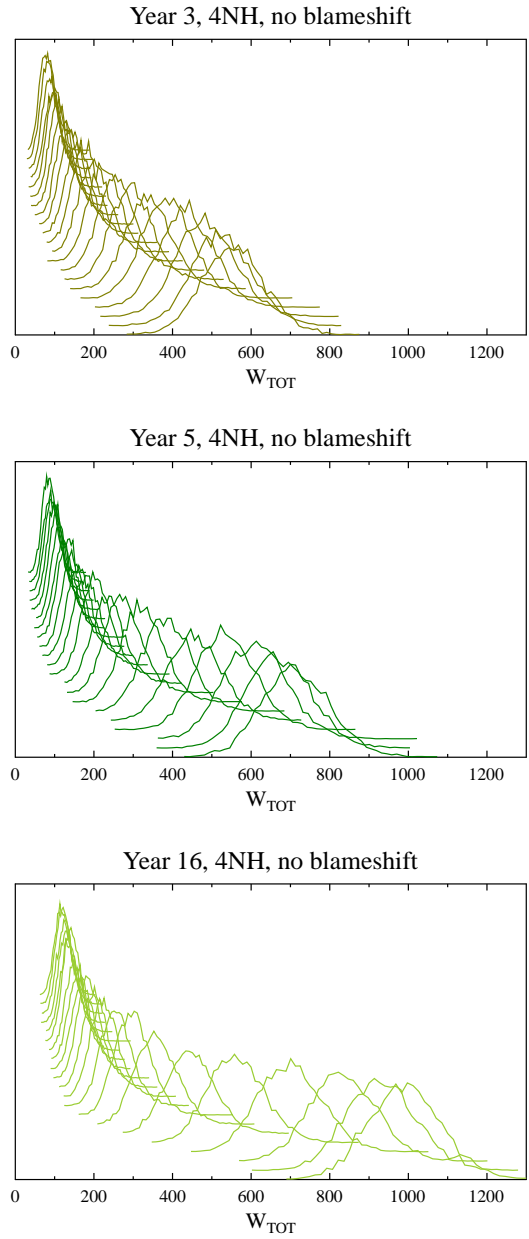}

\caption{Distribution of $W_{TOT}$ values for Peter model (organization with
5 levels with 5 positions in a workgroup) depending on $C$ value
after 3, 5 and 16 years. Probabilities of $W_{TOT}$ are very well
described by Gaussian distributions. Increased values of $C$ shift
the center of the distribution to smaller values and decrease the
width of the distribution -- top left curves in each panel correspond
to $C=5$, bottom right to $C=0.01$. \label{Flo:WTOTPet} The distributions
were displaced vertically for clarity for increasing $C$ values.}

\end{figure*}

\begin{figure*}
\includegraphics[scale=0.8]{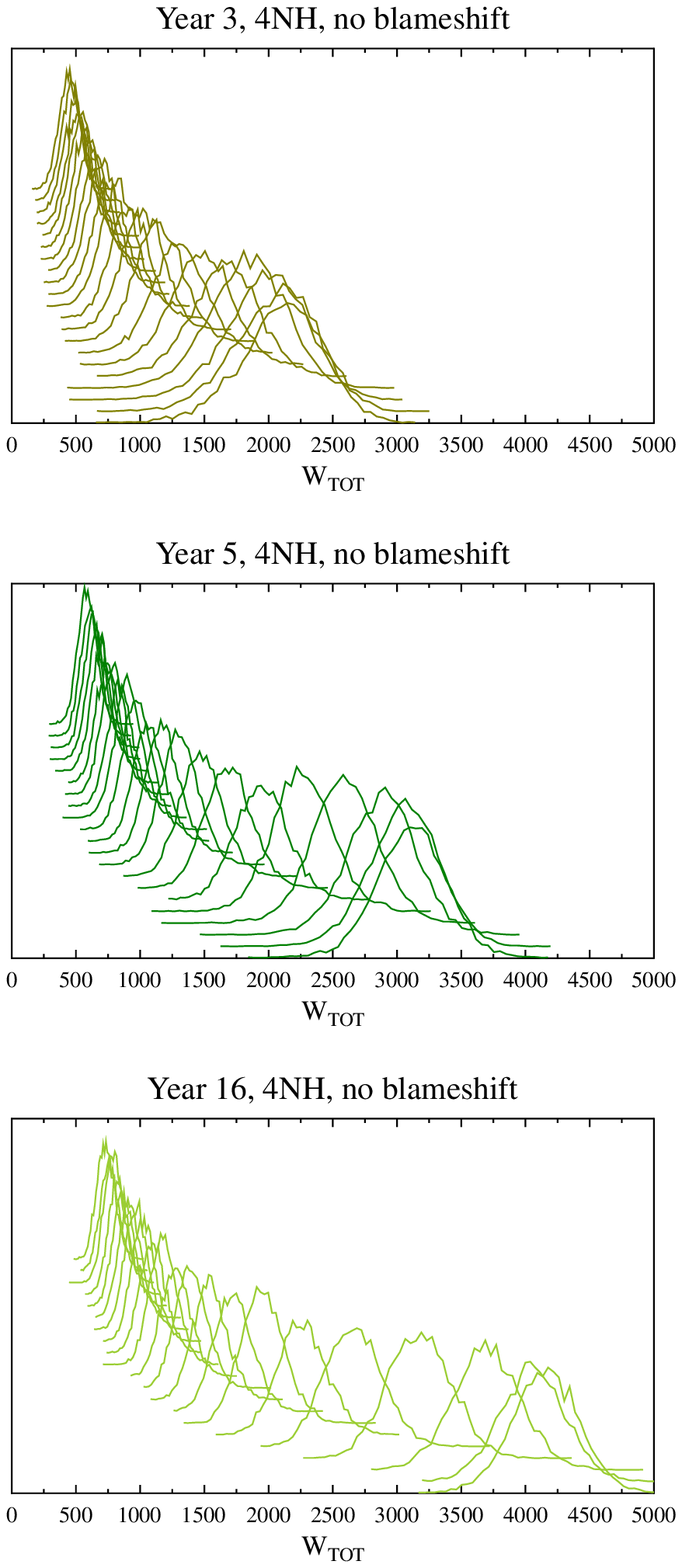}

\caption{Distribution of $W_{TOT}$ values for continuity model (organization
with 5 levels with 5 positions in a workgroup) depending on C value
after 3, 5 and 16 years. Probabilities of $W_{TOT}$ are very well
described by Gaussian distributions. \label{Flo:WTOTCon} The distributions
were displaced vertically for clarity for increasing $C$ values.}

\end{figure*}

Due to the multiplicative way the managerial contribution is modelled
here, the influence of individual characteristics of managers, especially
at high positions, can significantly change the overall productivity
of the organization. As a result we observe rather wide distribution
of results between simulation runs. As an example, Figure \ref{Flo:WTOTPet}
presents distribution functions of overall productivity $W_{TOT}$ for the Peter
model for various values of the susceptibility factor $C$,
at different stages of the evolution starting from random configuration: after3, 5 and 16 years. 
Figure \ref{Flo:WTOTCon} presents similar data for a set of simulations
in the continuity model.

\begin{figure*}
\includegraphics[scale=0.8]{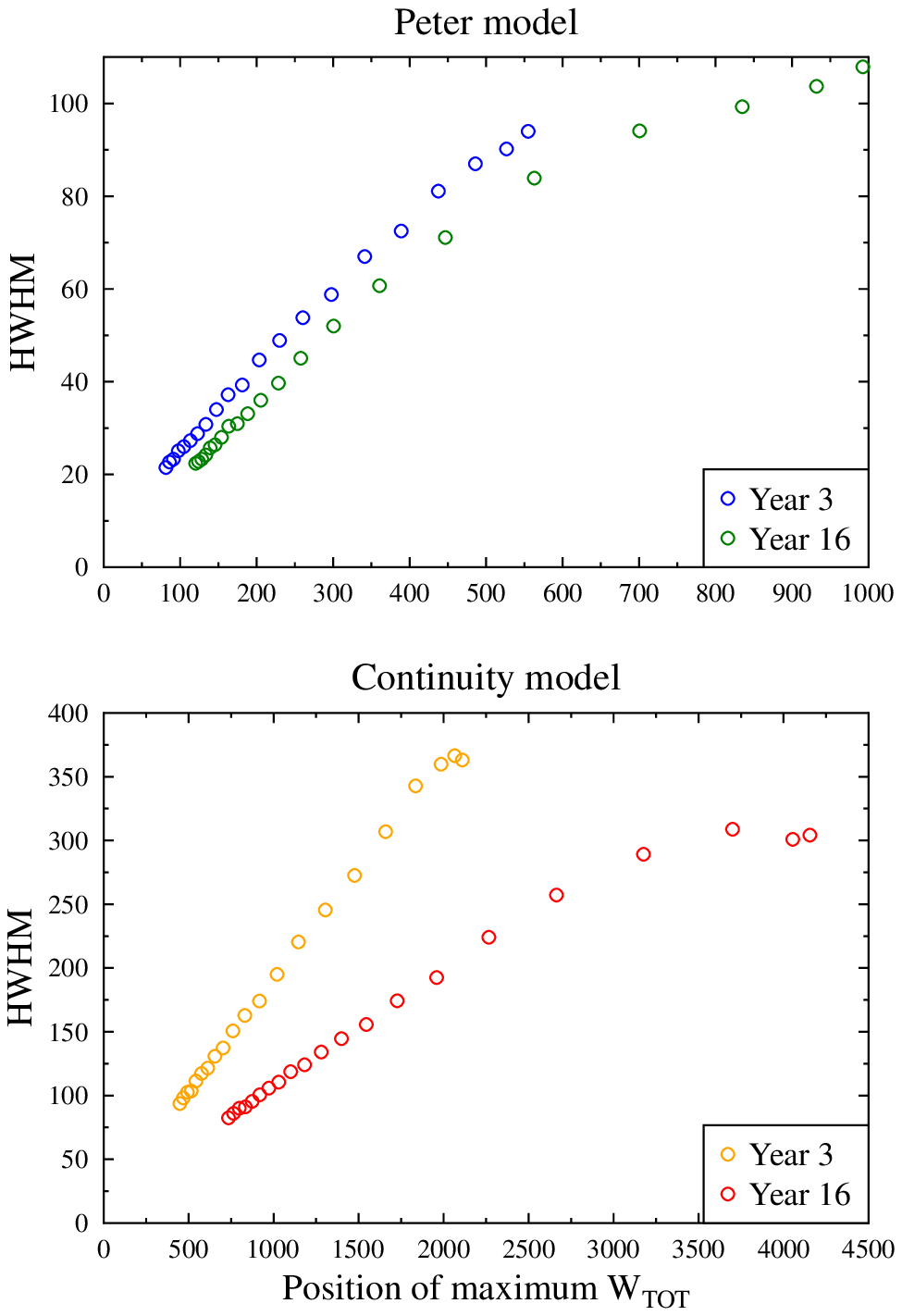}

\caption{Correlations between the maximum of the $W_{TOT}$ distribution in
multiple simulation runs and the width of the distribution. Simulations
for Peter and continuity models with 5 levels and 5 positions in a
workgroup, pre-selection between four candidates in external hiring
and no blame-shifting. Point series correspond to situation after
3 and 16 years, and each series is given by different susceptibility
$C$ values, ranging from $C=5$ at bottom left corner to $C=0.01$
at the top right corner. \label{Flo:width-peak}}

\end{figure*}

The widths of the distributions are rather high, with half-width at
half-maximum strongly correlated with the position of the peak center
(Fig. \ref{Flo:width-peak}). In the continuity model we observe strong
reduction of the width of the distribution of $W_{TOT}$ with the
passage of time, but even after 16 years of continuous evolution,
there are still significant differences of results between individual
simulation runs. Thus, part of the trends observed for average values
as functions of model parameters, discussed in the later part of this
article, may be masked by individual differences between simulation
runs. We note here that the average values of $\langle p_{i}\rangle_{k}$
and $\langle w_{i}\rangle_{k}$ for organization level $k$ have similar
Gaussian distributions. %

\subsection{Evolution of individual agent characteristics}

We note first that the behavior of the lowest and the mid-range hierarchy
levels is different. This is due to radically different mechanisms
of changes in the agents occupying these levels. The ratio of vacated
positions to level size was set as constant for levels below the topmost.
It should be noted, however, that the lowest level is filled entirely
from outside, which leads to strong dependence of the average characteristics
of agents on that level on the external hiring process -- especially
on the results of selection of the external candidates. 

\begin{figure*}
\includegraphics{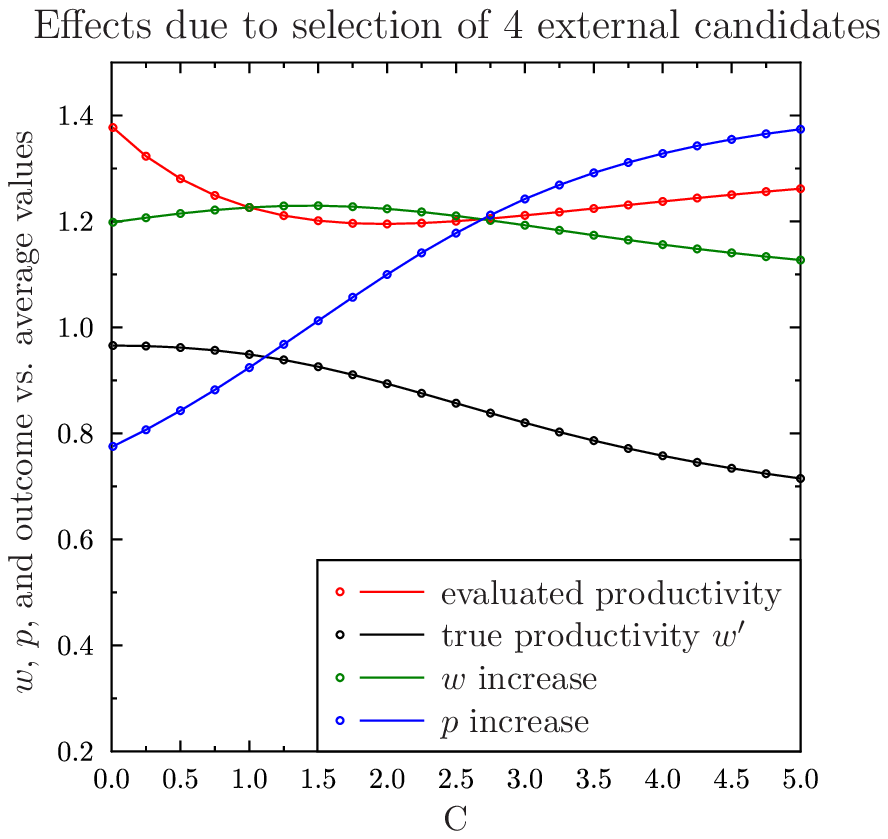}

\caption{Improvement of characteristics of externally hired agents due to pre-hiring
selection process, compared to simple random drawing of a single agent
characteristics. The red line denotes the individual perceived productivity, normalized to the average value,
$w_{i}'/(w_{0}-p_{max}/2)+C\cdot p_{i}$, which is the quantity evaluated and compared
during the pre-selection process. Due to different distributions of $w_{i}$
and $p_{i}$ the average values of these parameters behave differently.
For small values of $C$ the involvement in politics is selected against,
and resulting in decrease in $p_{i}$, while for high $C$ agents
with high values of political activity are preferred (blue line).
The black line shows the true average effective productivity $w_{i}'=w_{i}-p_{i}$
of the selected candidates, decreasing with the company susceptibility,
but still higher than the value of $w_{0}-p_{max}/2=0.7$ expected
for a single random new hire. \label{Flo:4nh1nh}}

\end{figure*}

\begin{figure*}
\includegraphics[scale=0.8]{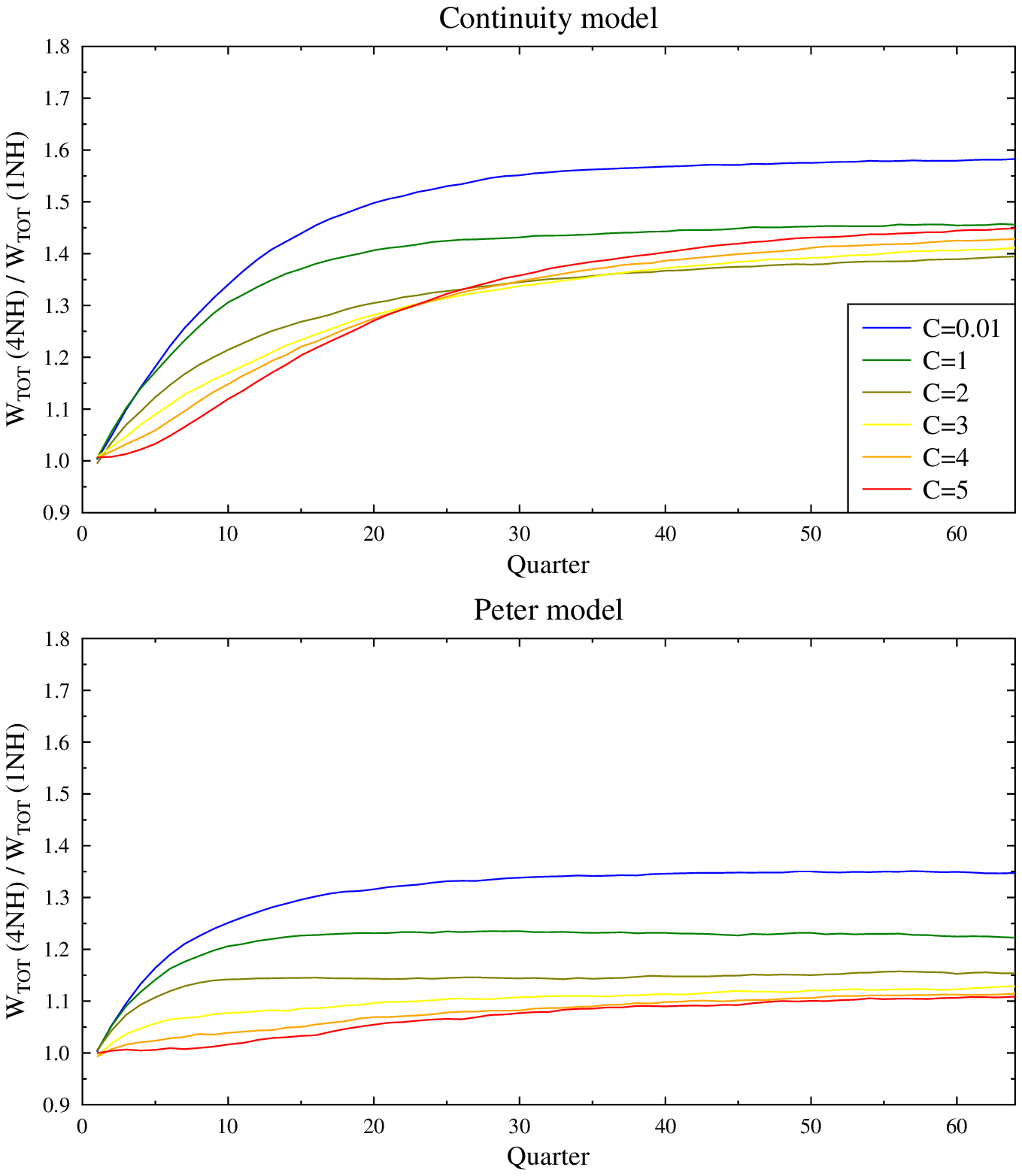}

\caption{Comparison of the overall organization performance $W_{TOT}$ evolution
with and without external hiring pre-selection. Upper panel - continuity
model, lower panel, Peter model. Lines for several values of company
susceptibility are shown. \label{Flo:wtotnh1nh}}

\end{figure*}

The most straightforward effects are due to the presence or absence
of selection of external candidates. We have compared results when
only one external candidate is randomly drawn (denoted 1NH) and when
the candidate with best perceived value of $w_{i}'/(w_{0}-p_{max}/2)+C\cdot p_{i}$
expected productivity is chosen from four candidates (denoted 4NH).
Without selection the expected values of $w_{i}$ and $p_{i}$ for
the newly hired employees would be $w_{0}$ and $p_{max}/2$, respectively.
As Fig. \ref{Flo:4nh1nh} shows, allowing the organization to choose from
only 4 candidates significantly influences the resulting values of
$w_{i}$ and $p_{i}$ and the resulting productivity. Such increase
in the real productivity $w_{i}'$ of newhires is especially important
for the lowest level of organization. For the Peter model, where only
$p_{i}$ are kept by an agent upon promotion, the increases of $w_{i}$
are largely lost at higher levels. For the continuity model both $p_{i}$
and $w_{i}$ are inherited and, as the new hires participate in later
promotion possibilities the values obtained with pre-selection on
higher levels also show an increase compared with simple random values
for external hiring. As a result, the overall performance of the organization
$W_{TOT}$ is higher in the 4NH case than in the 1NH case for both
Peter model and continuity model, but the difference is much higher
for the continuity model (50-70\% compared to 10-40\%, see Fig.\ref{Flo:wtotnh1nh}).

\begin{figure*}
\includegraphics[scale=0.8]{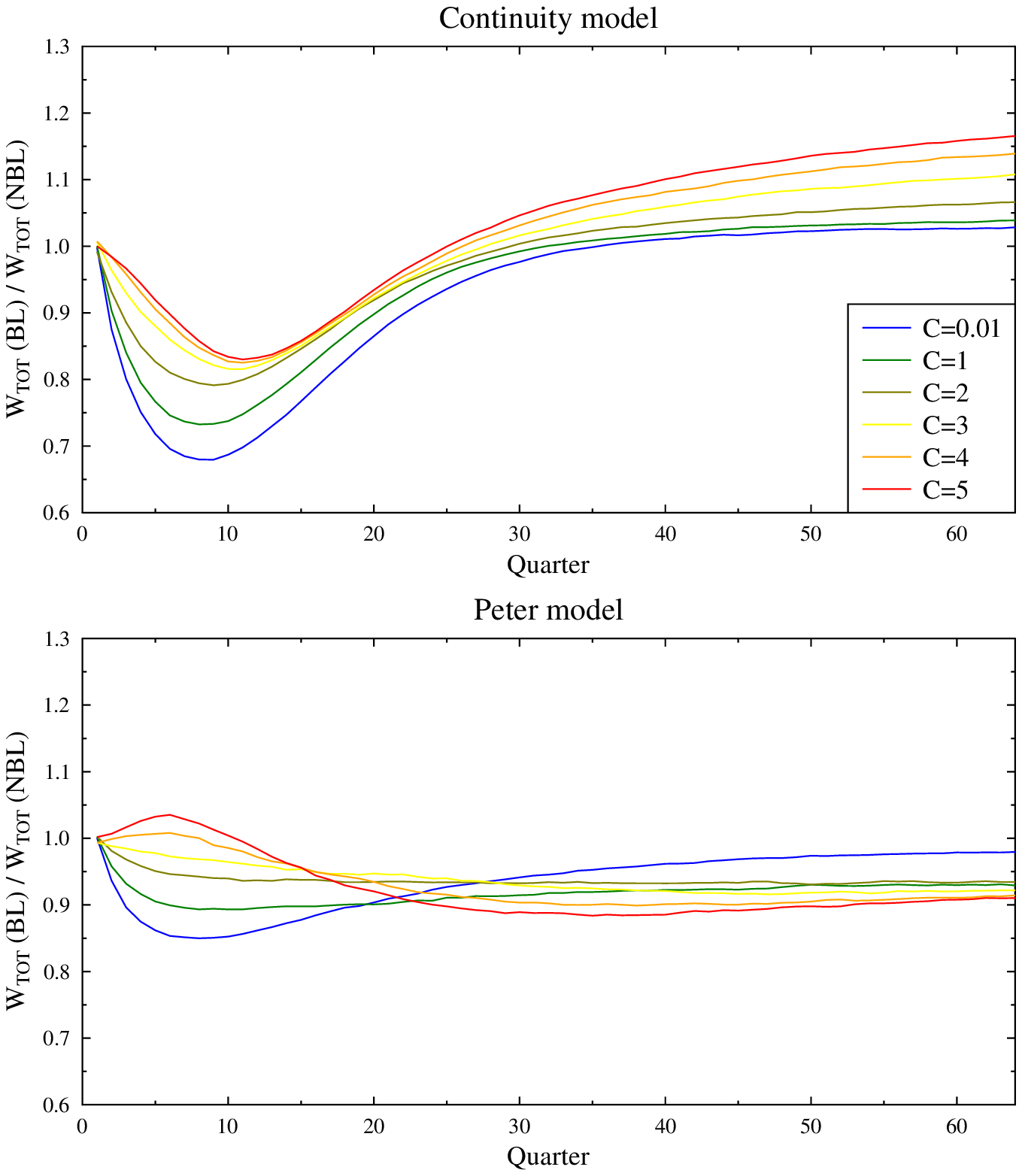}

\caption{Comparison of the overall organization performance $W_{TOT}$ evolution
with blame-shifting (BL) and without it (NBL). Upper panel - continuity
model, lower panel, Peter model. Lines for several values of company
susceptibility are shown. \label{Flo:wtotblnbl}}

\end{figure*}

The blame-shift process during which managers can place the responsibility
for the department weak performance on one of their subordinates acts
in a more subtle way. First, it extends the lifetime of managers,
especially the ones with weak real performance but high values of
$p_{i}$. Because we have assumed the probability of using a scapegoat
to be given by manager's $p_{i}$ (and independent on $C$) the effect
is the strongest for low values of $C$, where it leads to much larger
average values of $p_{i}$ at mid-managerial levels for small $C$.
This is especially visible early in the simulations, within the first
3 years of organization evolution from the random initial state. The
overall effects of the possibility to shift the blame for weak performance
are different for the continuity and Peter models (see Fig. \ref{Flo:wtotblnbl}).
For the Peter model there is a general but small decrease in organization
productivity when blame-shifting is possible. For the continuity model
there is a large decrease of productivity in initial stages that gets
gradually compensated at later times, so that for large $C$ values
the overall productivity in organizations allowing blame shifting
becomes even greater than for those that do not allow this process.

The time evolution of average values of $\langle w_{i}\rangle_{k}$
and $\langle p_{i}\rangle_{k}$ for various organization levels are
-- for most parameter combinations -- reasonably well described by
exponentially decaying functions of time $w(t)\approx w_{max}+(w_{0}-w_{max})\exp(-t/T_{w})$
and $p(t)\approx p_{max}+(p_{0}-p_{max})\exp(-t/T_{p})$. For very
low values of $C$ the self-promotion is selected against,
thus $p_{i}$ decreases with time for both continuity and Peter models.
For large values of $C>2$ selection favors high political activity
and thus $p_{i}$ grows. In the intermediate regime of $0.5<C<1.5$
we observe more complex behavior of $p_{i}$, remaining close to the
average value for random distribution. On the other hand, the true
productivity in both models is observed to grow with time. Obviously
this increase is much higher for the continuity model, where the promotion
preserves, at least partially, the $w_{i}$ values, so that talent and competencies are not lost. 

\begin{figure*}[h]
\includegraphics[scale=0.7]{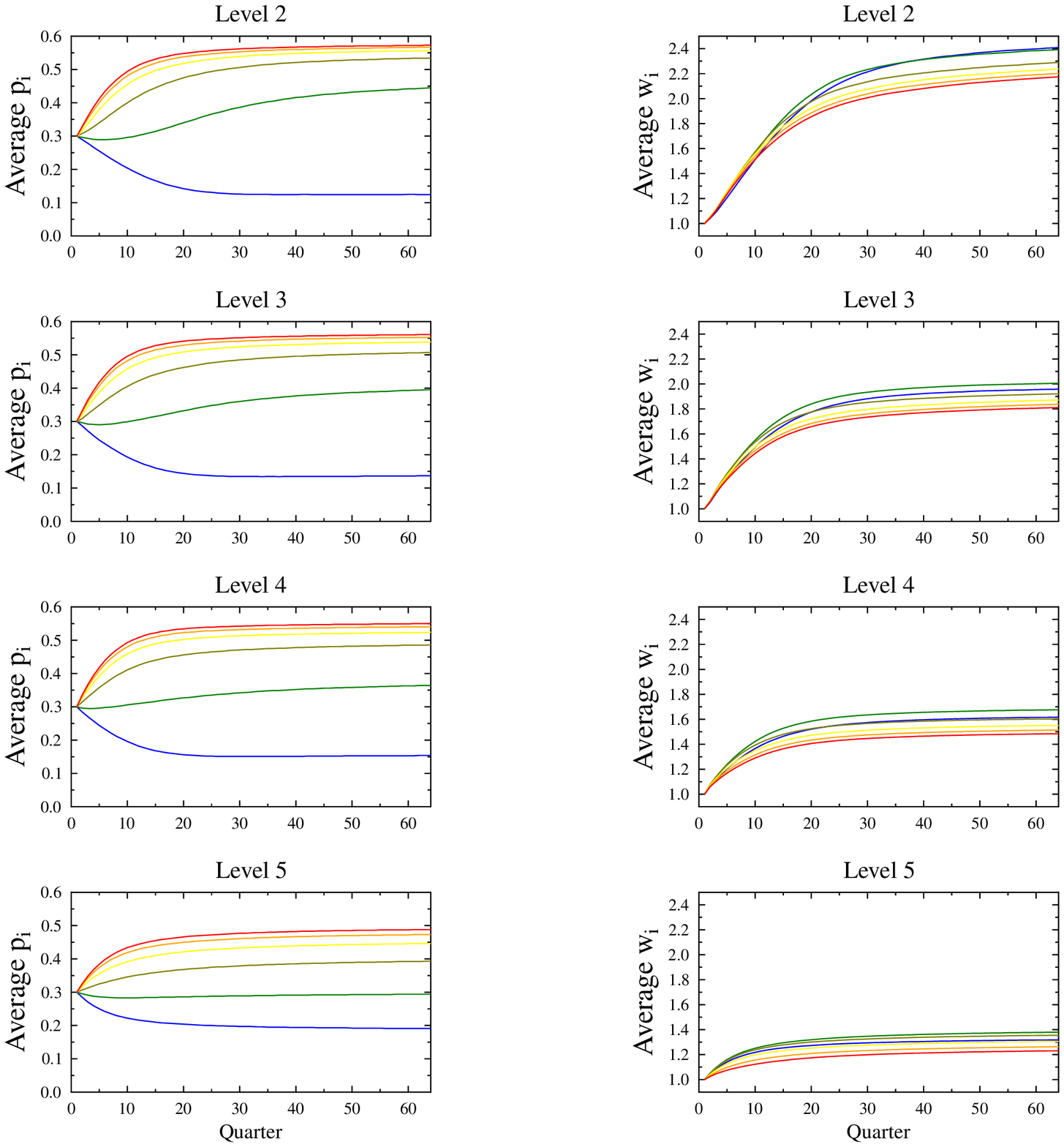}

\caption{Time evolution of average $\langle p_{i}\rangle_{k}$ and $\langle w_{i}\rangle_{k}$
values for the \textbf{continuity} model; 5 levels of 5 positions
in a workgroup, with blame-shifting and pre-selection of external
candidates (4NH).}

\end{figure*}

\begin{figure*}[h]
\includegraphics[scale=0.7]{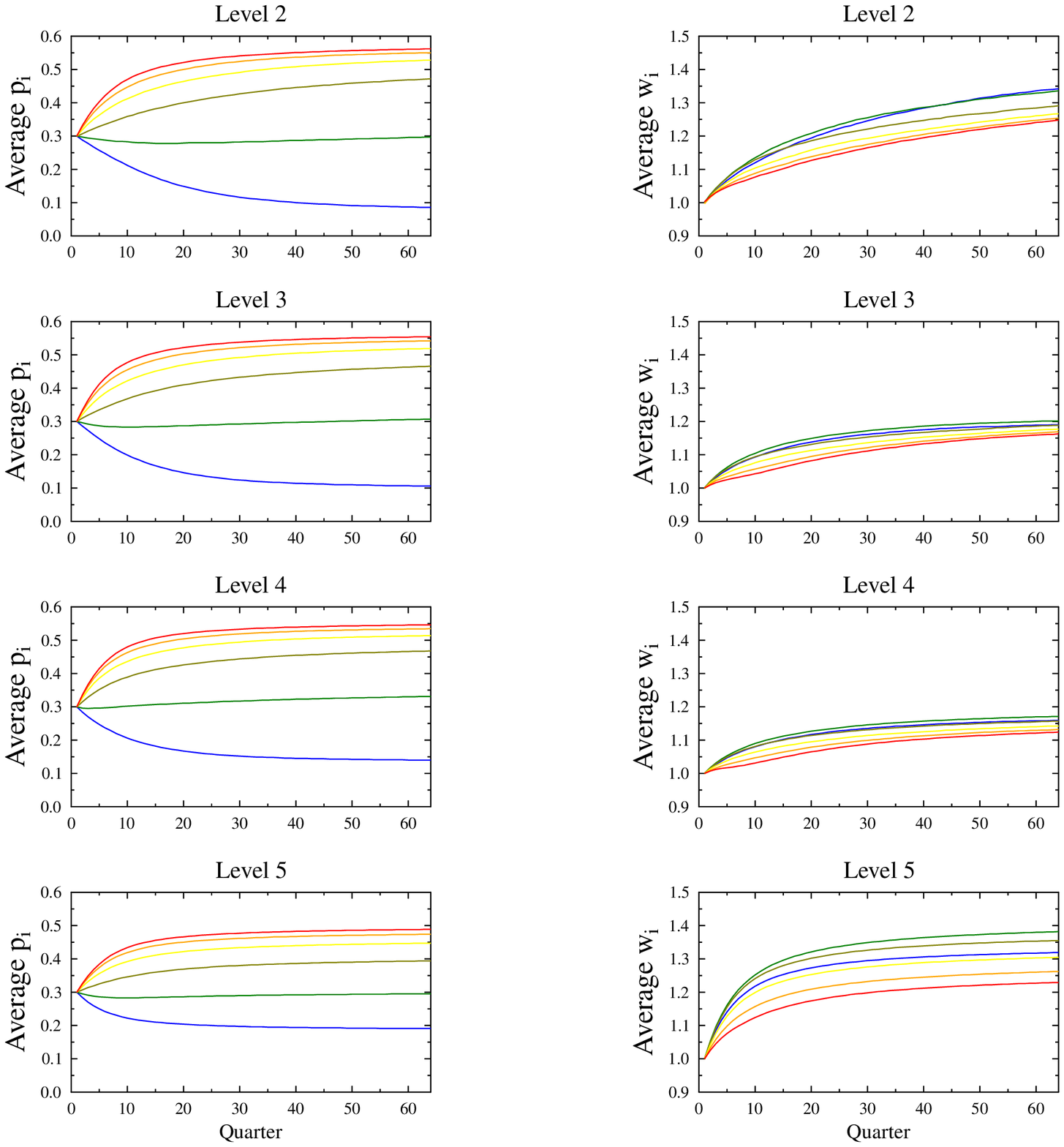}

\caption{Time evolution of average $\langle p_{i}\rangle_{k}$ and $\langle w_{i}\rangle_{k}$
 values for the\textbf{ Peter} model; 5 levels of 5 positions in a
workgroup, with blame-shifting and pre-selection of external candidates
(4NH).}

\end{figure*}

\begin{figure*}[h]
\includegraphics[scale=0.7]{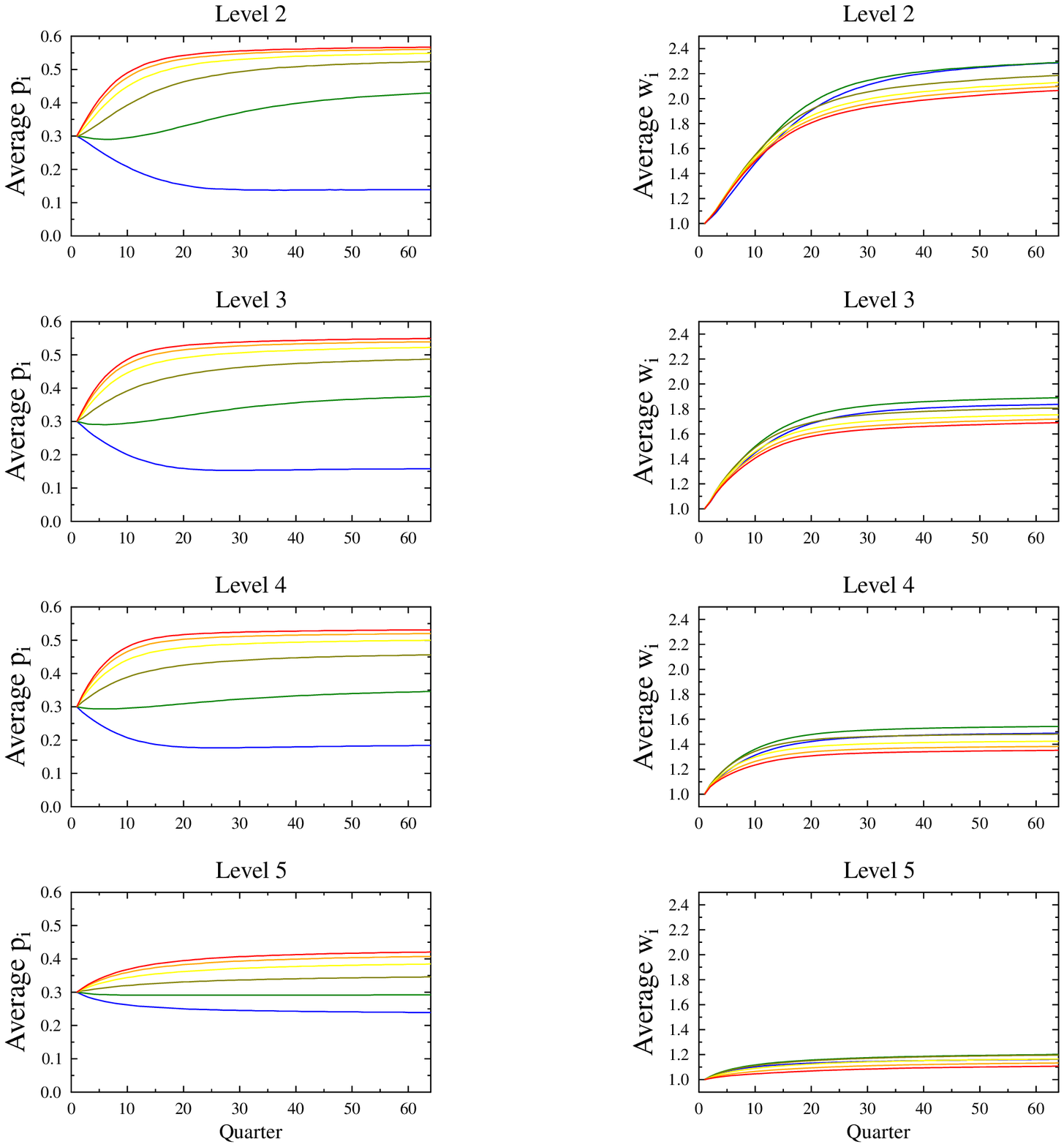}

\caption{Time evolution of average $\langle p_{i}\rangle_{k}$ and $\langle w_{i}\rangle_{k}$
values for the \textbf{continuity} model; 5 levels of 5 positions
in a workgroup, with blame-shifting and no pre-selection of external
candidates (1NH).}

\end{figure*}

\begin{figure*}[h]
\includegraphics[scale=0.7]{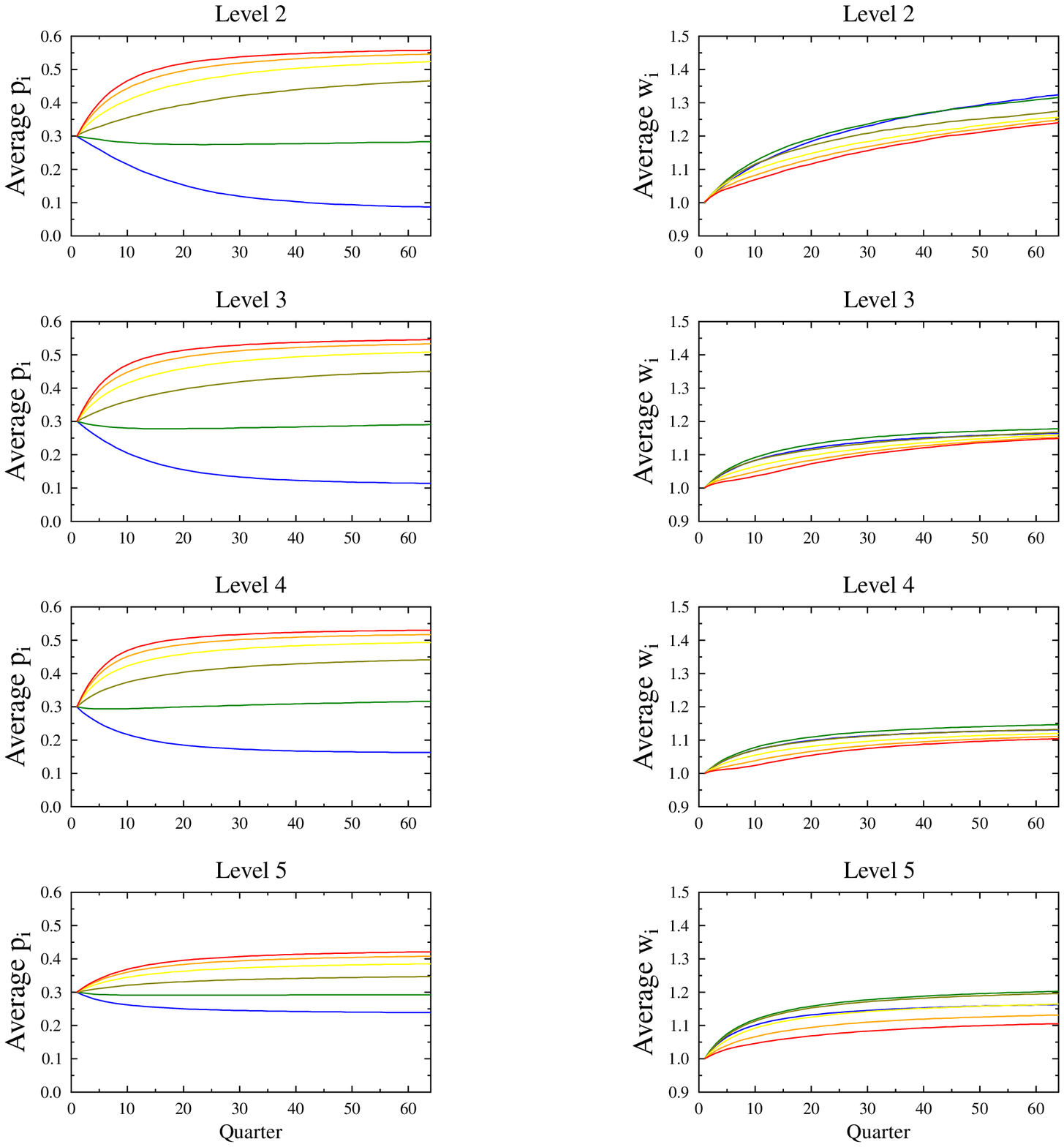}

\caption{Time evolution of average $\langle p_{i}\rangle_{k}$ and $\langle w_{i}\rangle_{k}$
values for the \textbf{Peter} model; 5 levels of 5 positions in a
workgroup, with blame-shifting and no pre-selection of external candidates
(1NH).}

\end{figure*}

\subsection{Global characteristics}

We shall focus now on the remaining model controls: the post-promotion
effectiveness model (Peter or continuity) and the organization susceptibility
to internal self promotion, $C$. Both factors play a crucial role
in the evolution of the organization effectiveness. Let us discuss
first the differences between the two post-promotion models. The continuity
model, even in the presence of strong susceptibility to internal PR,
predicts improvement of the individual productivity $w_{i}$ with
time at higher levels of organization. Selection of best (perceived)
performers coupled with limited {}``inheritance'' of the individual
productivity significantly improves the average values, and as result,
leads to much higher values of overall true productivity than in the
Peter model. 

\begin{figure*}[h]
\includegraphics[scale=0.8]{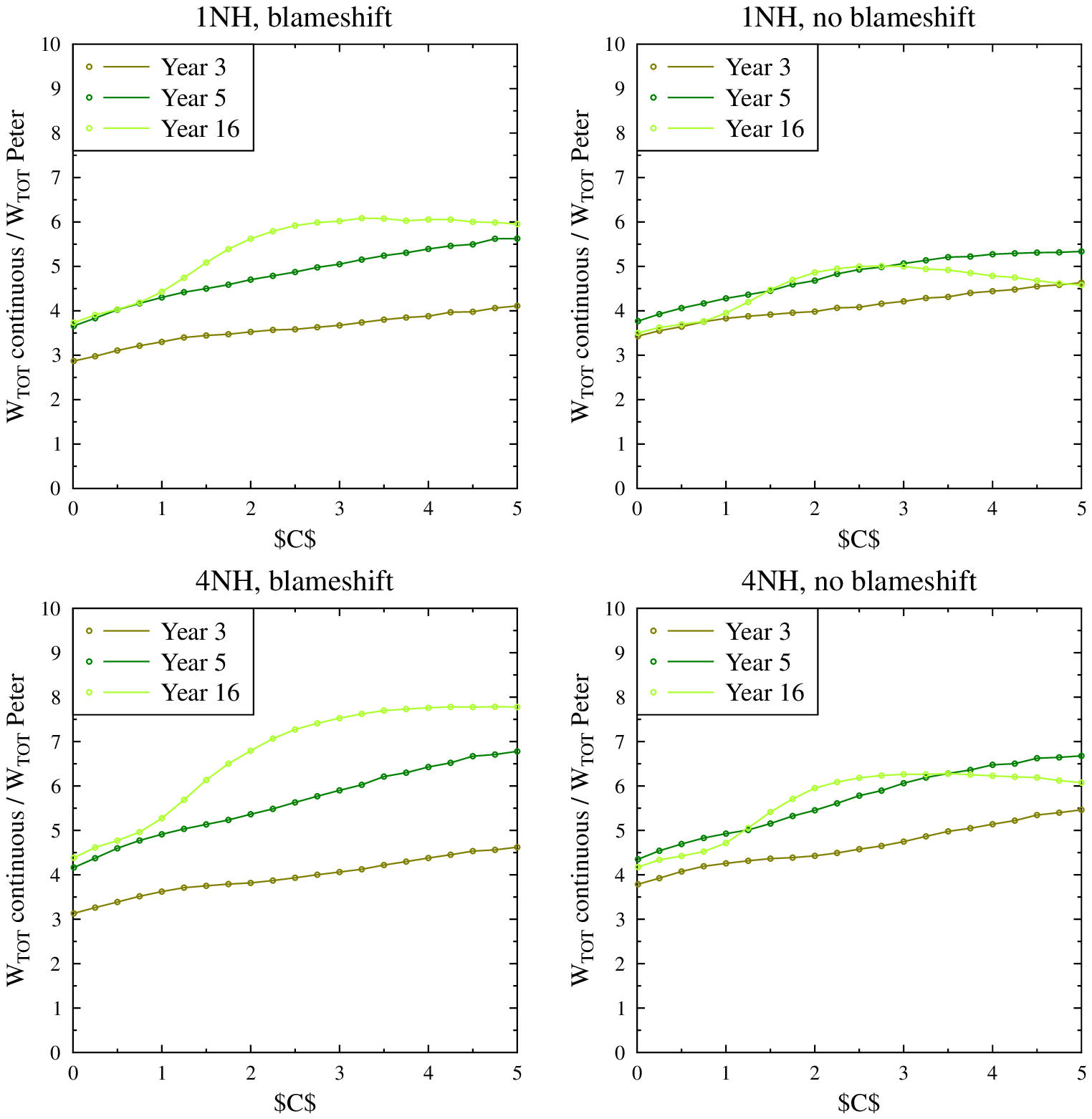}

\caption{Comparison of overall organization effectiveness for the continuity
and Peter models for various combinations of input parameters: with
or without blame-shifting and with or without pre-selection of newhires
(4NH and 1NH). Hierarchy with 5 levels and 5 positions in a workgroup
has been used in this simulation. \label{Flo:p-c-out}}

\end{figure*}

\begin{figure*}
\includegraphics[scale=0.8]{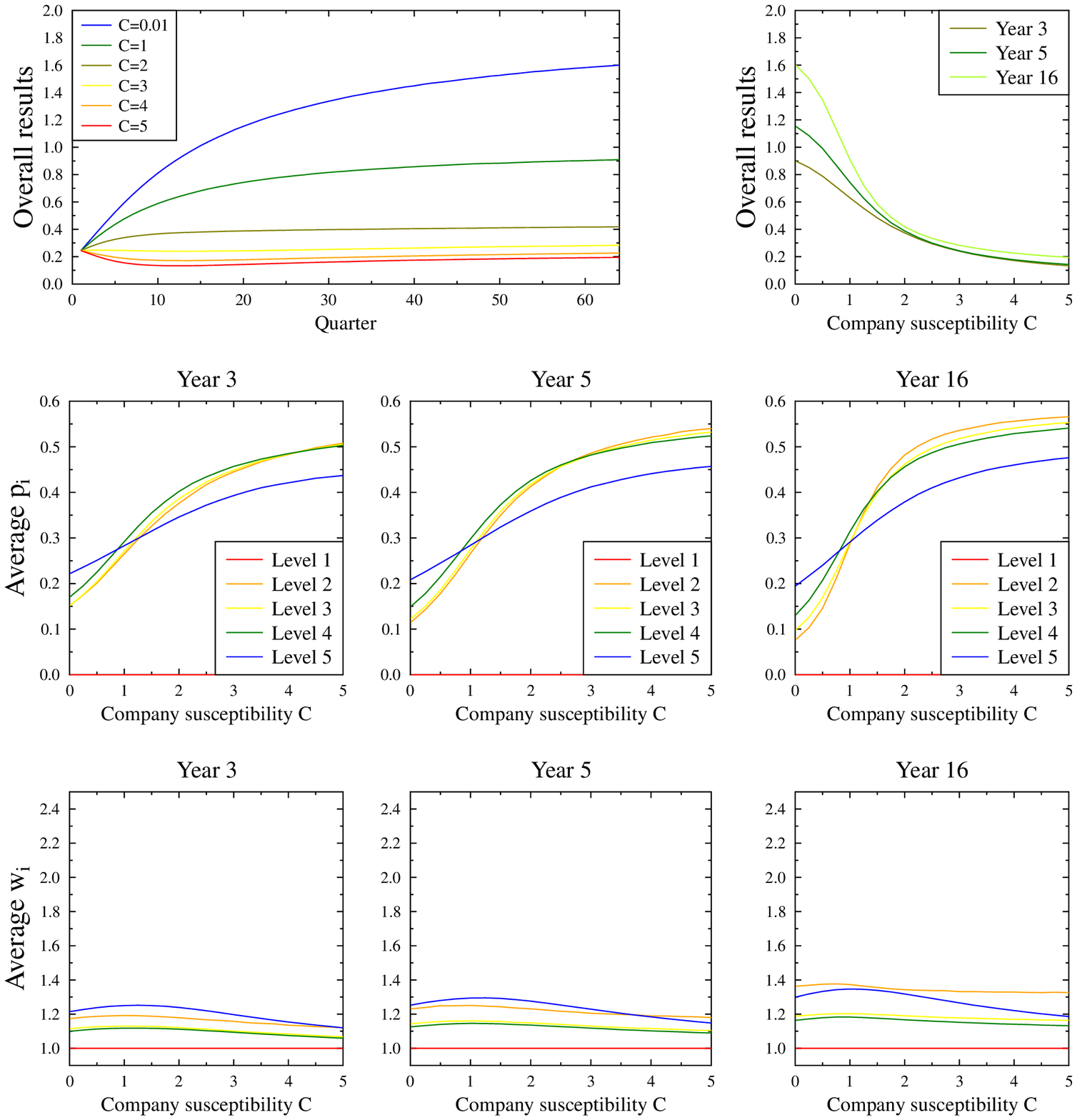}

\caption{Simulation results for the Peter model, 5 layers, 5 positions in a
workgroup. Pre-selection from 4 external candidates (4NH) and no no
blame-shift process present. Upper panels: time evolution and dependence
of overall productivity $W_{TOT}$ on organization susceptibility
$C$, divided by productivity of \textquotedbl{}neutral\textquotedbl{}
organization of the same size. Lower panels: dependence of average
$p_{i}$ and $w_{i}$ values on $C$ at the end of the third, fifth
and 16th year of evolution. \label{Flo:pet5x54nhbl-1}}

\end{figure*}

\begin{figure*}
\includegraphics[scale=0.8]{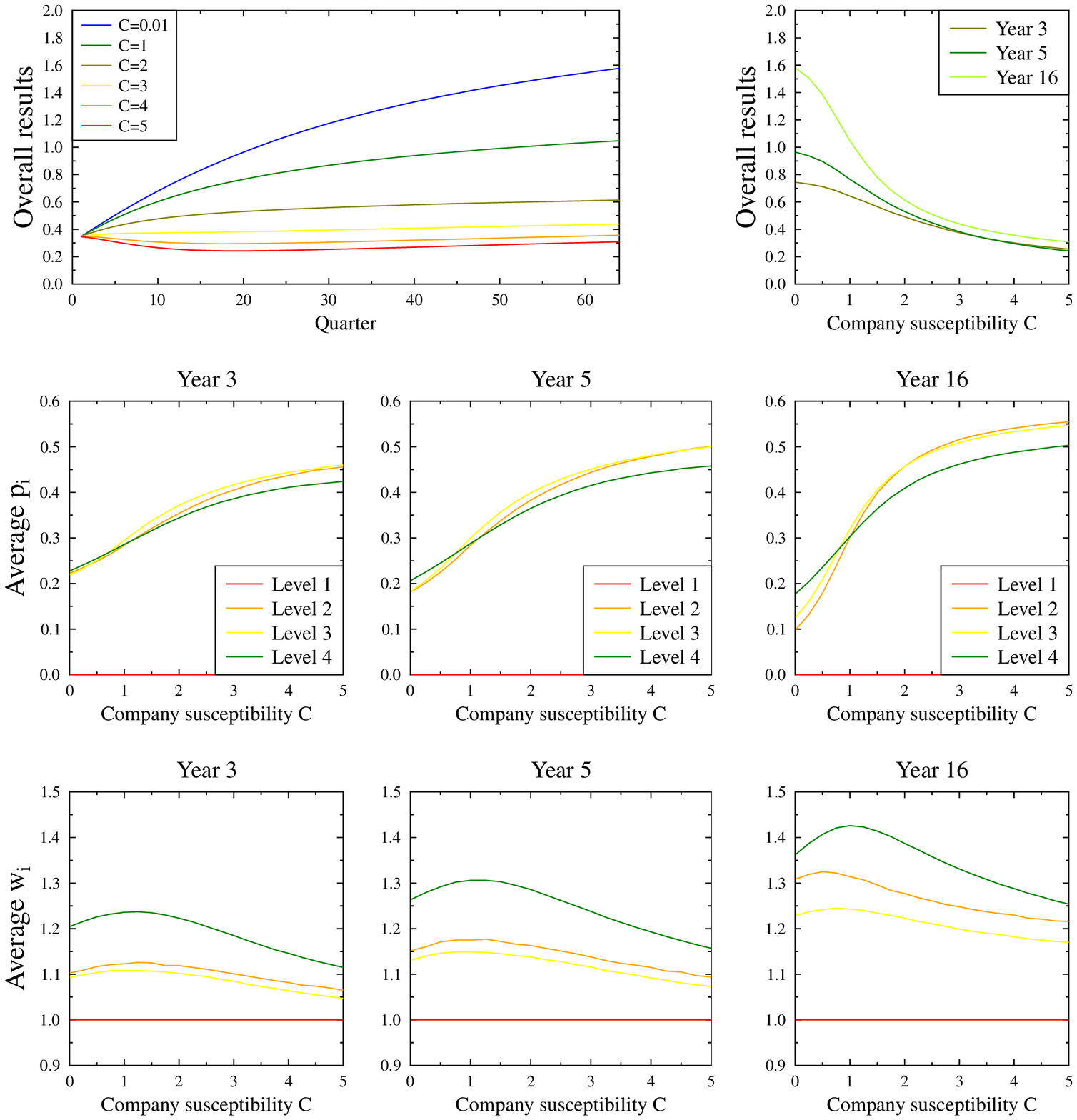}

\caption{Simulation results for the Peter model, 4 layers, 9 positions in a
workgroup. Pre-selection from 4 external candidates (4NH) and blame-shift
process present. Upper panels: time evolution and dependence of overall
productivity $W_{TOT}$ on organization susceptibility $C$, divided
by productivity of \textquotedbl{}neutral\textquotedbl{} organization
of the same size. Lower panels: dependence of average $p_{i}$ and
$w_{i}$ values on $C$ at the end of the third, fifth and 16th year
of evolution. \label{Flo:pet4x94nhbl-1}}

\end{figure*}

The difference between the productivity in continuity model and Peter
model can be as high as 8 times, as shown in Fig. \ref{Flo:p-c-out}.
Selection process for the continuity model,
even for very large values of susceptibility $C$ (when agents with high focus on self-promotion are at advantage), leads to performance
better than for the random assignment of agent capacities -- but lower
than the neutral configuration, when internal politics plays no role. 
For the Peter model at large $C$ values the evolution
may lead to \emph{decrease} of productivity from the starting
random configuration! This seems to be a very bad sign for any organization.
For some combinations of parameters this decrease, happening over
a short period of less than 10 quarters, diminished the productivity
by a factor of 2 (see, for example, Figs. \ref{Flo:pet4x94nhbl-1},\ref{Flo:pet5x54nhbl-1}).

\subsection{Conclusions}

The general results of the simulations are not surprising (just as
the original observations of Peter and Adams were not surprising).
Everyday observations show that there are many organizations, big
and small, commercial and governmental, where promotions and demotions
follow political ploys and not the real capability of
an employee to fulfill specific roles. And it is quite obvious to
both internal and external observers, that the performance of these
organizations may be far from optimal. Thus the model has no {}``discovery''
value, but is, more or less, a mathematical toy, reflecting
some aspects of the social reality. Obviously, it misses a
lot of factors that are present in real life: individuality and creativeness
of the leaders; innovative, market disruptive products or ideas; well established
processes and organizational culture that effectively guide individual players; capacity to change
the organizational form to adapt to new challenges. All these can
be crucial in determining the success or a failure of the organization.
On the other hand, the processes related to promotion and internal
politics included in the model are present in almost all types of
organizations and can negatively impact their results. The main idea
behind the model is to look for some simple controls that could allow
some general policy suggestions -- as it turns out some model parameters
are more important than others. For example, the presence of blame-shifting
influences the overall performance only to a minor degree. Moreover,
such self-preservation instincts are natural for managers, and it
s very hard to avoid them. Another model variable, degree of pre-screening
of external candidates, provides much higher influence on the resulting
productivity (especially in the continuity model). But for most of
modern organizations such competitive nature of hiring is already
present, so there is little room for improvement here. The two major
factors are the capability to preserve the skills and efficiency of
an employee after the promotion and susceptibility of an organization
to self-promotion, or, in other words, capacity to recognize the real values. 
Here the differences in overall productivity can
be as high as an order of magnitude. Thus, the model suggests that
organizations should focus on measures eliminating negative effects
due to Peter Principle and to self-promotion. Such measures could
include:
\begin{itemize}
\item Using measurable and objective criteria for employee evaluation. This
is relatively easy for some areas (for example in sales departments),
but rather difficult in creative environments (scientific research, software engineering).
Yet despite the difficulty, such standards would decrease the employee's
drive to use political skills for self-promotion, instead of focusing
on the needed tasks.
\item Giving prospective candidates for promotion  tasks
related to the nature of duties at the higher level (e.g. temporary management of
small groups, responsibility for analysis of results and preparing directions for action for workgroups and departments 
etc.) and measuring results of such assignments. These results should be
used when considering promotion, as they estimate the productivity
at the higher level, and thus decrease the effect of Peter Principle. 
\item Introduction of horizontal advancement paths, in which employees who
do not fit into the traditional promotion model (for example brilliant
engineers lacking managerial skills) would still be able to achieve
satisfaction within a company, without falling prey to the Peter Principle
trap.
\end{itemize}
Even a moderate decrease of the company susceptibility to political
ploys, and improvement of the {}``heritability'' of skills after
promotion may lead to dramatic improvement of overall productivity,
not by single percentage points but by a significant factor. 

\subsection{Model extensions}
The initial model presented here can be expanded in several directions.
The best source of improvement of the model would be when the computer
simulation could be coupled with some {}``microscopic'' sociology
studies (for example interviews focused on measuring the payoff of
political activities within the organization). Especially, if one
could provide comparative studies of general effectiveness of organizations
built upon different social models. 

The model itself is quite flexible and allows many improvements. Such
extensions of the computer model would still miss the effects due
to individuality of participants and specific nature of the organization
but bring the simulated strictures still closer to reality. For example
it would be interesting to study results of changes in organization
policies related to promotion and susceptibility to self-interests
in a {}``mature'' environment. In such simulations an organization
resulting from some years of evolution under one set of parameters
would be used as the starting point of a new simulation, for example
with a changed value of $C$. This would correspond to corrective measures
undertaken by top management, especially after a major change. The
main question of such studies would be to determine, for example,
the expected improvements due to lowering of susceptibility or the
time it takes to see such improvement take the effect. 

To make the model more realistic one might divide the organization into
a few `divisions'. Promotion within a division (e.g. sales, marketing,
manufacturing) should be based on a common sense hypothesis, as a
lot of workload remains the same, so the effective results should
be changing only partially. On the other hand, promotion across the
departments should be less frequent and the new value of raw productivity
would be totally uncorrelated. 

Yet another
direction of further research is the study of dependence of overall
output and effectiveness per employee in a growing organization --
as most of real life organizations are dynamically changing their
size.

The model can also be improved  by including
effects of employee dissatisfaction and restlessness when they are
not promoted. In the current paper, the external candidates
come from an infinite pool of agents with random characteristics.
In real life they would be the people from appropriate levels of other
organizations (usually quite similar to the studied one). By symmetry,
this effects should be included in the study: a process of agents leaving
the company to join other ones. In contrast with firing of the worst
perceived performers, here the agents most likely to leave would be
the ones who have relatively high opinion of themselves and willingness
to seek new opportunities. These qualities, in the first approximation
are related to the self-promotion drive $p_{i}.$ The probability
should increase with time spent at the same position. As a result,
increases of productivity introduced by the pre-screening of external
candidates would be counterbalanced by the outflow of the restless,
disgruntled employees. 

\onecolumngrid

\bibliographystyle{unsrtnat}

\cleardoublepage

\subsection*{Appendix: Detailed simulation results}

In this section we present more detailed simulation results for several
sets of model parameters. We focus on the time evolution of average
organization productivity (starting from a random configuration),
dependence of this productivity on the susceptibility $C$ after 3,
5 and 15 years of evolution and similar dependence of the average
$p_{i}$ and $w_{i}$ values at various levels of the organization.
Overall productivity has been normalized to that of a {}``neutral''
organization, in which all employees have neutral productivity ($w_i=1$),
and no self-promotion ($p_i=0$), which means that the managers neither
increase nor decrease their team output. In such neutral configuration,
the productivity is given by the sum of work done by the workers at
the lowest level. Such normalization allows to see the effects independently
to organization size and number of levels.

\end{document}